\documentclass{iopart}
\usepackage{iopams}
\usepackage[ bookmarks, colorlinks=true, plainpages = false, citecolor = red, urlcolor = blue, filecolor = blue]{hyperref}
\usepackage{graphicx}
\usepackage{epstopdf}
\usepackage{bm, bbm}
\usepackage{wasysym}
\usepackage{cite}
\usepackage{mathrsfs}

\newcommand{\be}{\begin{equation}}
\newcommand{\ee}{\end{equation}}
\newcommand{\nn}{\\ \nonumber}
\newcommand{\bra}{\left\langle}
\newcommand{\ket}{\right\rangle}

\def\a{\alpha}
\def\b{\beta}
\def\k{\kappa}
\def\l{\lambda}
\def\ve{\varepsilon}

\newcommand{\I}{\mathrm{i}}
\newcommand{\E}{\mathrm{e}}

\newcommand{\M}{\mathscr{M}}
\newcommand{\Imod}{\mathscr{I}}
\newcommand{\Vmod}{\mathscr{V}}
\newcommand{\Lmod}{\mathscr{L}}
\newcommand{\K}{K}

\newcommand{\bos}{\varphi}
\renewcommand{\IP}{\mathcal{I}}

\newcommand{\op}{\phi}
\newcommand{\pop}{\partial\phi}
\newcommand{\opa}{\upsilon}
\newcommand{\poplog}{\psi}
\newcommand{\Tlog}{\tau}

\newcommand{\idsub}{1}

\newcommand{\opasub}{\upsilon}
\newcommand{\poplogsub}{\psi}
\newcommand{\Tlogsub}{\tau}

\newcommand{\f}[1]{\phi(x_{#1})}
\newcommand{\vp}[1]{\mathcal{V}_\phi^+(x_{#1})}
\newcommand{\vn}[1]{\mathcal{V}_\phi^-(x_{#1})}

\newcommand{\scp}[1]{\mathcal{V}_+(x_{#1})}
\newcommand{\scn}[1]{\mathcal{V}_-(x_{#1})}
\newcommand{\ICI}{$\left\{\op\op\right\}_1$}
\newcommand{\STI}{$\left\{\op\op \right\}_T$}

\newcommand{\lsep}{\cdot}
\newcommand{\lsept}{\hspace{-1pt}\cdot\hspace{-1pt}}
\newcommand{\p}{\partial}
\newcommand{\cs}{\left| \chi \right\rangle}
\newcommand{\xs}{\left| \xi \right\rangle}
\newcommand{\keto}[1]{\left| #1\right\rangle}

\newcommand{\Eq}{{\cal E}}
\newcommand{\F}{{\cal F}}

\newcommand{\Do}[1]{{\cal D}_{#1}}
\newcommand{\Del}[2]{\Delta_{#1}^{#2}}

\newcommand{\Doa}[1]{\widetilde{{\cal D}}_{#1}}
\newcommand{\Dela}[2]{\widetilde{\Delta}_{#1}^{#2}}

\newcommand{\Dob}[1]{\widehat{{\cal D}}_{#1}}
\newcommand{\Delb}[2]{\widehat{\Delta}_{#1}^{#2}}

\begin{document}

\title{Logarithmic operator intervals in the boundary theory of critical percolation}
\author{Jacob J~H~Simmons}
\address{Maine Maritime Academy, Pleasant Street, Castine, ME, 04420}
\eads{\mailto{jacob.simmons@mma.edu}}

\date{\today}

\begin{abstract}
We consider the sub-sector of the $c=0$ logarithmic conformal field theory (LCFT) generated by the boundary condition changing (bcc) operator in two dimensional critical percolation.  This operator is the zero weight Kac operator $\phi_{1,2}=:\op$, identified with the growing hull of the SLE$_6$ process.

We identify  percolation configurations with the significant operators in the theory. We consider operators from the first four bcc operator fusions: the identity and $\op$; the stress tensor and it's logarithmic partner; $\pop$ and it's logarithmic partner; and the pre-logarithmic operator $\phi_{1,3}$.

We construct several intervals in the percolation model, each associated to one of the LCFT operators we consider, allowing us to calculate crossing probabilities and expectation values of crossing cluster numbers.  We review the Coulomb gas, which we use as a method of calculating these quantities when the number of bcc operator makes a direct solution to the system of differential equations intractable. 

Finally we discuss the case of the six-point correlation function, which applies to crossing probabilities between the sides of a conformal hexagon.  Specifically we introduce an integral result that allows one to identify the probability that a single percolation cluster touches three alternating sides a hexagon with free boundaries.  We give results of the numerical integration for the case of a regular hexagon.
\end{abstract}


\maketitle

\section{\label{sec: intro}
Introduction}

There is a long history of using conformal symmetry and boundary conformal field theory (CFT) methods \cite{BPZ,Cardy84,Dotsenko1984,BYB, Gurarie1993,Smirnov2001} to probe the problems of critical two dimensional percolation.  Successful predictions include a wide variety of crossing probabilities and expectation values for cluster numbers \cite{LanglandsEtAl92,Cardy92,Watts1996,Cardy2000,SimmonsKlebanZiffJPA07,SKFZ2011} and the applicability of these methods is now seldom questioned.  However the $c=0$ theory is a highly non-trivial logarithmic conformal field theory (LCFT) and it still remains to identify the role played by the various parameters and operators in the theory in order to build a cohesive $c=0$ theory applicable to both the bulk and boundary sectors \cite{Gurarie1999,Gurarie2005,MathieuRidout07,Rassmussen2007,Dubail2010, Vasseur2011,Vasseur2012,Vasseur 2012a,Ridout2012,Runkel2012}.

In this paper we examine the role of regular and logarithmic operators in the $c=0$ boundary conformal field theory of two dimensional critical percolation.  Our goal is to describe a correspondence between the basic operators of the theory and the percolation observables they represent. We'll describe the relation between fusion channel and boundary conditions, contrast logarithmic operators with their regular partners, and explain why logarithmic terms are relatively rare in the conformally invariant observables we normally consider.

In section \ref{sec: O n Loop} we review the $O(n)$ loop model \cite{Domany1981,Nienhuis1984,FortuinKasteleyn1972,DiFrancesco1998}.  This model is well suited for a representation where conformal blocks correspond to distinct boundary conditions.   The $O(n)$ loop model is closely related to the more rigorous multiple Schramm--L\"owner Evolution or Conformal Loop Ensembles (SLE$_\k$/CLE$_\k$) \cite{Schramm2000_UST_LERW,LawlerSchrammWerner01,BauerBernard2003,BauerBernardKytola05,Dubedat2006,Kytola09}.  We discuss the relation between these models only briefly.

In section \ref{sec: VertexOp}, we'll quickly review the Coulomb Gas formalism \cite{Nienhuis1984,Kondev1997}.  We introduce this formalism primarily for convenience as it allows us to write integral expressions for multi-point quantities that are otherwise difficult to solve. It is easy to identify these integral expressions with conformal blocks, which allows us to write properly normalized integral expressions for the correlation functions that we need.   A recent article by Flores and Kleban addresses the issues of whether these integral expression are exact solutions to the relevant 2$N$ point functions \cite{FloresKleban2012} with favorable results.  We will beg the question through out.

In section \ref{sec:Perc} we consider critical percolation.  We'll focus on identifying the percolation cluster configurations that are associated with the operators that appear in combinations of up to four boundary change operators.  By constructing intervals that fuse to these operators we'll be able to associate each operator with a percolation configuration.

We'll consider operators from one pre-logarithmic and two staggered logarithmic modules.  This collection of modules has seven significant operators.  The highest order operators in the two staggered modules are the identity and the boundary change operator, which both have simple interpretations.   We offer interval constructions for the five remaining operators and give Coulomb gas constructions for each.

Focusing on the two regular/logarithmic operator pairs, we find that the regular operator is a \emph{passive} operator while the logarithmic partner is an \emph{active} operator. Passive operators feel the effect of percolation clusters that have already been marked by other operators in the correlation function.  Active operators mark additional clusters; increasing the complexity of observables described by the correlation function.

In section \ref{sec: 3 Interval Xing} we'll use these identifications, and the Coulomb gas constructions to derive integral expressions for crossing probabilities between three disjoint boundary intervals; we call this geometry a conformal hexagon. We discuss several distinct results in the conformal hexagon including new crossing probabilities in hexagons with homogeneous free boundaries on all sides.

In \ref{Appendix} we present a multi-point operator product expansion for three $\phi_{1,2}$.  These multi-point expansions can be derived with minimal assumptions about the LCFT of the boundary theory: that the highest weight operator is primary, that it obeys the standard second order differential equation, and that it's non-null first order descendant has a logarithmic partner.

\section{The $O(n)$ loop model \label{sec: O n Loop}   }

In this section we describe the basic fusion channel boundary condition correspondence for a more general model then just critical percolation. We'll use the $O(n)$ loop model, of which percolation is a specific case corresponding to the  $n=1$ dense phase.  The discussion in this section should also apply to any systems described by conformal loop ensemble or Schramm--Loewner evolutions.  In section \ref{sec:Perc}, when we return specifically to the percolation model, we'll discuss how to modify this correspondence to include logarithmic operators in the $c=0$ LCFT.  

The $O(n)$ loop model \cite{Domany1981} is a generalization of the usual $O(n)$ spin model, governed by the truncated high temperature partition function
\be
Z_{\Omega} = \mathrm{Tr} \prod_{\langle i j \rangle}\left( 1+x\, \boldsymbol{s}(r_i) \cdot \boldsymbol{s}(r_j)\right)\; ,
\ee
with $n-$component spins $\boldsymbol{s}(r_i)$ with squared norm $n$ on each site of a lattice $\Omega$.
We expand the product and associate a graph on $\Omega$ to each term by including the bond between  $r_i$ and $r_j$ if the $x\, \boldsymbol{s}(r_i) \cdot \boldsymbol{s}(r_j)$ factor appears in the term and excluding the bond if it does not.  Now the on-site trace for an odd number of spin components is zero, so the only graphs that contribute are those composed entirely of closed loops.  

The form of the resulting loop partition function is particularly simple if we choose the honeycomb lattice, $HC$, for $\Omega$.  In this case the loops can visit each site a maximum of one time and the per site trace, $\mathrm{Tr} s_a(r_i)s_b(r_i) = \delta_{a b}$, means that each loop earns a net weight $n$.  Combining this with the weight $x$ per occupied bond yields the partition function
\be \label{eq:part_func}
Z_{HC}=\sum_{\Lambda} n^{N_\ell} x^L\; ,
\ee
where the sum is over all graphs of closed loops $\Lambda \subset HC$, and $N_\ell$ is the number of loops, and $L$ the total length of the loops in each configuration. Taking (\ref{eq:part_func}) as our starting point, the model generalizes immediately to non-integer $n$.

For small values of $x$, long loops are suppressed and the model flows to the vacuum under renormalization.  For large $x$, long are favored and the system flows to a fixed point of densely packed loops under renormalization.  For $|n|<2$, the boundary between these two regimes is $x_c=(2+\sqrt{2-n})^{-1/2}$, representing the dilute critical point \cite{Nienhuis1984}.

\subsection{\label{sec: Arc Config}
Arc Configurations and Boundary CFT}

\begin{figure}[htbp] 
\centering
\includegraphics*[width=0.5\columnwidth]{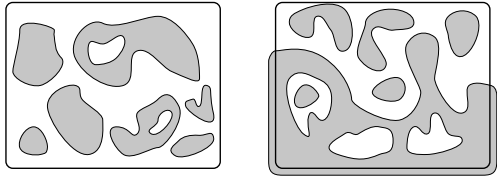} 
\caption{\label{fig:On_v_SLE}
LEFT: a schematic of an ${\rm O}(n)$ configuration without boundary operators, when all loops lie in the interior of the region. RIGHT: an illustration of a configuration with an , which effectively travels outside of the region in order to close.
}
\end{figure}

We consider the boundary theory generated by $\phi_{1,2}$ Kac operators.  For simplicity we'll write $\op:=\phi_{1,2}$.  These operators represent  points where the ends of open loop, which we call \emph{arcs}, are anchored to the boundary.  We will sometimes refer to $\op$ as a 1-leg operator, with the understanding that 2 legs will close in the bulk to form 1 arc. In the spin version of the $O(n)$ model, anchoring a loop from $r_a$ to $r_b$ on the boundary is accomplished by inserting the factor $\boldsymbol{s}(r_a)\cdot \boldsymbol{s}(r_b)$ into the partition function.  In the loop model the equivalent operator is an external loop segment running from $r_a$ to $r_b$, as in \fref{fig:On_v_SLE}.

In the first part of this paper we focus on correlation functions involving $N$ arcs,
\begin{equation} \label{equ:2N_Correlation_function}
\bra \op(x_1)\op(x_2)\op(x_3)\ldots\op(x_{2N-1})\op(x_{2N})\ket\; .
\end{equation}
The convention for closing the open bulk arcs fixes the weights of each configuration and we call the corresponding external loops \emph{boundary conditions}. These boundary conditions correspond to conformal blocks in the associated CFT, and are computationally tractable for large numbers of boundary operators, largely through the Coulomb gas formalism, which we review in section \ref{sec: VertexOp}.

It can be argued (in some cases rigorously \cite{Smirnov2001, Smirnov2010}) that non-trivial fixed points of the $O(n)$ loop model correspond to Schramm--L\"owner Evolution or Conformal Loop Ensembles with parameter $\k$ (SLE$_\k$/CLE$_\k$) related to $n$ by
\be \label{n_of_kappa}
n=-2 \cos\left(4 \pi/\k\right)\;,\quad \left\{ \begin{array}{ll}(2<\k<4)\quad&\textrm{dilute}\\( 4<\k)\quad&\textrm{dense}\end{array} \right.\; .
\ee
Thus \eref{equ:2N_Correlation_function} represents a $2N$--SLE process where the legs grow stochastically with a time like parameter until they pair off to form arcs.  These functions were studied by Dub\'edat \cite{Dubedat2006} using SLE commutation relations to generate differential equations equivalent to those found using CFT. Currently, we are interested in the final arc configurations of these processes, which are naturally described in language of the $O(n)$ model.

With $2N$ boundary operators the associated loop configurations each have $N$ non-crossing arcs connecting the operators pairwise. There are $(2N)!/N!(N+1)!$ distinct pairings \cite{DiFrancesco1998} and we segregate loop configurations into classes depending on the final arc configuration.  We denote the set of all such classes by $\Omega$.
Once we've decomposed the configurations into arc classes, we define \emph{arc weights} by restricting the partition sum to a given class and giving a fugacity $n_{\mathrm{arc}}=1$ to each arc.  This allows us to rewrite the partition function as a sum over arc classes:
\begin{equation*}
Z_{BC}=\sum_{\omega \in \Omega} F_{BC}^\omega W_\omega\;,\qquad \textrm{where} \qquad W_{\omega}=\sum_{[\Lambda] = \omega} n^{N_\ell} x^L\; .
\end{equation*}
These arc weights form a natural basis for the $2N$--\,point correlation functions.  The extra factors in the partition function are determined purely by the interaction of bulk arcs and the boundary conditions we assign.

In CFT we calculate conformal blocks, not arc weights.  Conformal blocks are defined through a sequence of operator \emph{fusions}. Fusion identifies two or more neighboring operators with an operator series located at a single point.  The contribution of a single CFT module to the fusion product defines a fusion \emph{channel}.  When we specify an order to fuse the operators and the resulting channels for a correlation function we define a conformal block.  Often in statistical mechanics, different conformal blocks contribute depending on the physical meaning we assign to the operators.  We'll introduce a physical boundary condition/conformal block correspondence that allows us to determine $F_{BC}^\omega$.

Throughout this article we adopt notation from \cite{MathieuRidout07} for the modules we encounter.  Unquotiented Verma modules are denoted by $\Vmod$, while the irreducible quotient of these modules are denoted by $\Lmod$.  We use $\M$ for modules when only specific singular descendants are to be null.

Weights in the body of the Kac table occur in pairs $h_{r,s}=h_{r',s'}$ so that the maximal submodules of $\Vmod_{r,s}$ are generated by singular operators at levels $rs$ and $r's'$.  In this case $\M_{r,s}$ is the quotient of $\Vmod_{r,s}$ by the null submodule at level $rs$.  This module is reducible, since it contains the null submodule at level $r's'$, but it is indecomposable.

Weights from the edge of the Kac table lead to modules with one maximal singular submodule.  In these cases we generally observe that $\M_{r,s}=\Lmod_{r,s}$.

Finally, staggered modules $\Imod_{r,s}$ are generated by a logarithmic operator $\phi_{r,s}$ that couples to a non-null singular descendant of some primary $\phi_{r',s'}$.  This module has a highest order maximal submodule isomorphic to $\M_{r',s'}$ and the quotient $\Imod_{r,s} \backslash \M_{r',s'}$ is isomorphic to $\M_{r,s}$ which represents the non-null descendants of $\phi_{r,s}$.  For brevity, we'll often use the operator symbol in place of Kac indices, \emph{i.e} $\M_{\phi}:=\M_{1,2}$.

Contractions in \eref{equ:2N_Correlation_function} are dominated by the fusion $\op \times \op =\M_1\oplus \M_\opasub$, with two products: the identity module and the module of the two leg operator $\phi_{1,3}=:\opa$. Fusion is a local property, and boundary conditions we associate with a certain fusion channel will connect the two fusing operators.  The difference in the two channels comes from what we allow the arcs attached to these boundary conditions to do. For two adjacent $\op$'s there are two case: either a single arc attaches them to each other, or separate arcs attach to them to another set of operators.  We call these cases \emph{contractible} or \emph{propagating} arcs respectively, to describe their limiting behavior as the interval between the two $\op$ shrinks to a point.

\begin{figure}[htbp] 
\centering
\makebox[0.1 \columnwidth]{\raisebox{2ex}{$\{ \op \op \}_1$:}} \includegraphics*[width=0.275\columnwidth]{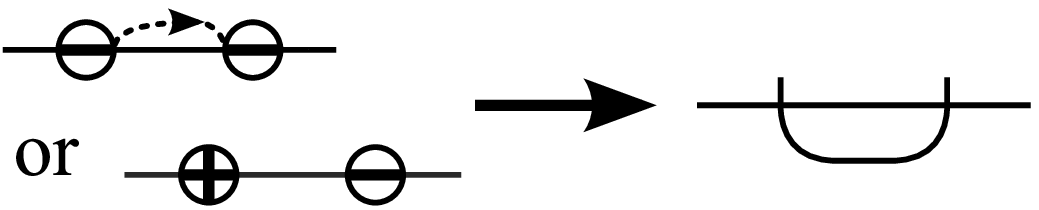}
\hspace{1cm}
\makebox[0.1 \columnwidth]{\raisebox{2ex}{$\{ \op \op \}_\opa$ :}} 
\includegraphics*[width=0.275\columnwidth]{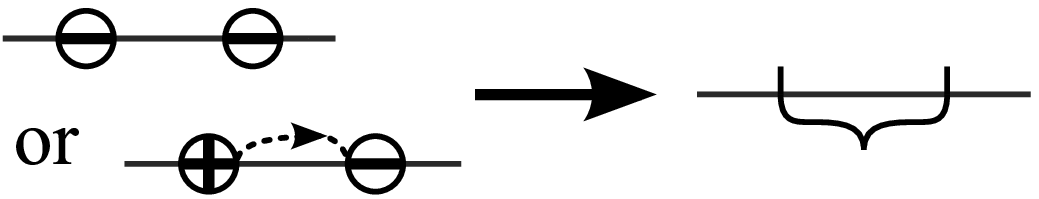}
   \caption{ \label{fig:11_Convention}
Fusion channel, Coulomb gas, and boundary conditions.
LEFT (Identity): Boundary condition shows how to close arcs in the partition function.
RIGHT (Two-leg): The bracket forbids contractible arcs between these points.}
\end{figure}

For the identity fusion product $\phi \times \phi = \M_\idsub$ it's natural to place no restrictions on the connectivities of the arcs, and the fusion should allow both contractible or propagating arcs. The corresponding boundary condition is an external segment connecting the two operators.   When two adjacent operators exhibit this fusion, we call the resulting interval an identity channel interval.  We denote and identity channel interval between $x_a$ and $x_b$ by $\{ \op(x_b) \op(x_a)\}_\idsub$.  See \fref{fig:11_Convention}  for the operator, boundary condition, and Coulomb Gas notation.

The two-leg fusion product $\phi \times \phi=\M_\opasub$ appears when we prohibit the operators from sharing a contractible arc. We indicate this prohibition, by connecting these operators with an external bracket instead of a smooth loop.  A loop containing a single bracket gets a weight zero, but a loop containing multiple brackets gets the regular weight $n$.  We call intervals that uniquely exhibit this fusion two-leg channel intervals, denoted  by $\{ \op(x_1) \op(x_2)\}_\opasub$.
See \fref{fig:11_Convention}  for the operator, boundary condition, and Coulomb Gas notation together.

For the purpose of assembling correlation functions of many boundary change events the Coulomb gas (CG) formalism is invaluable, since it allows us to write expressions for correlation functions with very specific fusion products, and thus very specific boundary conditions. Next we will show how to distinguish between the intervals that represent the $\op \times \op =\M_1$ and $\op \times \op = \M_\opa$ fusion channels in the CG formalism.

\subsection{Coulomb gas representations \label{sec: VertexOp}}

In this article we'll discuss 6--point functions for which no closed solutions have yet been found, so the primary analytic tool we'll use is the Coulomb gas (CG) formalism: a mapping from a loop model to a height model that leads to a transparent bosonization of the corresponding conformal field theory.

The structure of the Coulomb gas is far richer than the CFTs it is so often used to describe, and it is not the goal of this paper to establish a correspondence  between the operator content of the two theories.  Instead we use Coulomb gas techniques to identify integral expressions for correlation functions of type (\ref{equ:2N_Correlation_function}).  This identification requires that the integral expressions we construct obey the $2N$ second order differential equations implied by the $\phi_{1,2}$ null state.  Recently Flores and Kleban wrote on this rigorous connection between CFT and the Coulomb gas  in the case of (\ref{equ:2N_Correlation_function}), and we refer interested reader to their work \cite{FloresKleban2012,FloresKleban2013}.

First, we expand the configuration space of the system by giving a direction to each loop. Vertices gets a complex local weight $\exp(i \chi \theta / 2)$, where $\theta$ is the angle each loop turns though in traversing the vertex.  A closed loop thus receives a total weight of $\exp(\pm i \chi \pi)$ depending on whether it is oriented clockwise or counter-clockwise. The trace over each loop orientation produces a weight $n=2 \cos(\chi\, \pi)$ and we recover the $O(n)$ weights \cite{Nienhuis1984}.

Furthermore, directed loops are equivalent to the level lines of a restricted solid-on-solid height variable living on the lattice faces: the height increases (decreases) by $\pi$ whenever we cross a loop that points to the left (resp.~right).  The power of the CG formalism lies in the assumption that the continuum limit of this height model is a Gaussian free field. This gives us a bosonic field theory representation of the loop gas CFT.  

The bosonic field $\bos$ has action $S=S_O+S_C+S_D$, a standard Gaussian action $S_O$ to which we add two additional terms:
\begin{equation*} \fl
S_O = \frac{g}{4 \pi}\int (\nabla \bos)^2 \mathrm{d}^2x \; , \quad S_C = \frac{\I \a_0}{8 \pi} \int {\cal R} \bos\, \mathrm{d}^2x\; ,\quad \textrm{and} \quad S_D= a \int \cos 2 \bos\, \mathrm{d}^2x \; .
\end{equation*}
The field is coupled to the local Gaussian curvature ${\cal R}$ by the term $S_C$, in analogy with the weights $\exp(i \chi \theta/2)$. Loops in regions with non-zero curvature wind through angles $ \theta \neq 2 \pi$ and adding $S_C$ compensates for this (non-local) effect.
The stress tensor is altered by $S_C$, modifying the central charge: $c=1-24 \alpha_0{}^2$.

The discrete values of the restricted height model suggest the presence of non-irrelevant locking term $S_D$.  The long distance action remains Gaussian only if this term is marginal, which fixes the CG coupling constant  in terms of the loop fugacity and SLE parameter, $g=1-\chi=4/\k$ \cite{Kondev1997 
}. This correspondence implies that these loop models are described by CFTs with central charge and Kac weights
\be \label{KacWeights}
c=\frac{(3g-2)(3-2g)}{g} \quad {\rm and} \quad h_{r,s}=\frac{(r-s g)^2-(1-g)^2}{4 g}\; .
\ee

The operators in the theory are represented by vertices
\begin{equation*}
\mathcal{V}_{\alpha}(z) = \mathrm{e}^{\mathrm{i} \sqrt{2} \alpha \bos(z)}\; ,
\end{equation*}
where $\alpha$ denotes the vertex charge.  With the plain bosonic action, correlation functions with a non-zero charge vanish.  However, $S_C$ induces an anomalous background charge, and charge neutrality requires $\sum_i \alpha_i = 2\alpha_0=g^{-1/2}-g^{1/2}$ instead.  The form of a charge neutral correlation function is determined by the Gaussian action:
\begin{equation} \label{equ:charge_neutral_correlation}
\langle \prod_i \mathcal{V}_{\alpha_i}(z_i)\rangle =\prod_{j>i} (z_j-z_i)^{2\alpha_j \alpha_i}\; .
\end{equation}
Two point correlation functions are non-trivial when $\alpha_2=2 \alpha_0-\alpha_1$.  Comparing this with a CFT two point function
\begin{equation*}
\langle \mathcal{V}_{\alpha}(x) \mathcal{V}_{2\alpha_0-\alpha}(0)\rangle=\langle \phi_h(x) \phi_h(0) \rangle=x^{-2h}
\end{equation*}
shows that scaling operators can be represented by vertices with charges, $\alpha$ or $2 \alpha_0-\alpha$, which are related to the conformal weight by $h=\alpha(\alpha-2 \alpha_0)$.  The charges associated to the Kac operators, $\phi_{r,s}$ with weights given by (\ref{KacWeights}), are 
\be
\alpha_{r,s}^{\pm}=\frac{1\pm r}{2} \alpha_+ + \frac{1\pm s}{2} \alpha_-\; .
\ee
We call the two corresponding choices of vertex operators $\mathcal{V}_{r,s}^\pm$.  For brevity, we'll often use the operator symbol to replace the Kac indices such as $\mathcal{V}_{\phi}^\pm:=\mathcal{V}_{1,2}^\pm$.

For correlation functions of three or more operators, charge neutrality can significantly limit the set of computable correlation functions.  We can expand this set by including screening charges: non-local zero weight charges that modify the total charge of the correlation function without changing the local scaling properties of the vertex operators.  Screening charges are formed by weight one vertices integrated along non-contractible paths, $\IP$. The two possible screening charges are
\begin{equation*}
Q_{\pm}^{\IP} = \oint_{\IP} \mathrm{d}z \mathcal{V}_{\pm}(z)\; ,
\end{equation*}
where $\mathcal{V}_\pm$ denotes a vertex with charge $\alpha_{\pm}=\pm g^{\mp1/2}$, the solutions to $1 = \alpha(\alpha-2 \alpha_0)$.  
\begin{figure}[htbp] 
\centering
\includegraphics*[width=0.15\columnwidth]{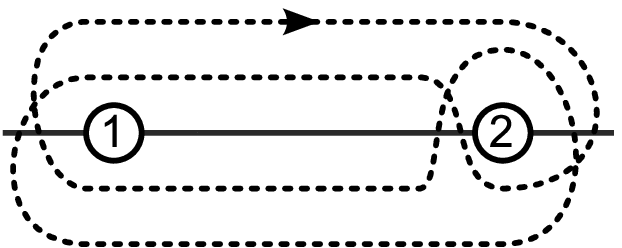}\;
\makebox[0.26 \columnwidth]{\raisebox{1.8ex}{$=4 \sin \b^\pm_1 \pi\; \sin \b^\pm_2 \pi$}}\;
\includegraphics*[width=0.15\columnwidth]{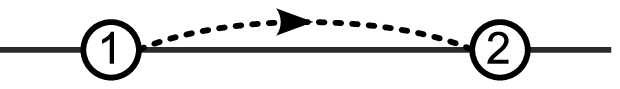}
   \caption{The dashed line on the left is a typical screening charge integration path entwining two charges $\a_1$ and $\a_2$.   If $\b^\pm_i:=2 \a_{i} \a_\pm>-1$ for $i=1,2$, then the integration can be equivalently performed along the real line.}
   \label{fig:CvR_Contour}
\end{figure}
In certain cases it's possible to replace the integration path $\IP$ with an equivalent real integration path, as in \fref{fig:CvR_Contour}, though we'll always draw these paths as real integrals for simplicity.  This proves to be an important point for $c=0$, and we'll discuss it in \sref{pop}.

Whenever a closed curve can be drawn around two operators without intersecting any of the $\IP$, we can identify the conformal highest weight of the pair's fusion product.   To see why, take \eref{equ:charge_neutral_correlation} in the limit where the curve and all of the vertices in its interior, including screening charges, are taken far away from the remaining operators, \emph{i.e} see the equations preceding \eref{equ:K_derivation}.  The leading order contribution is proportional to a correlation function that replaces the interior vertices with a single vertex operator.  The charge of this operator is equal to the sum of all the interior charges allowing us to identify the highest weight of the fusion product. The exponents of higher order contributions all differ by integers, which is consistent with contributions from descendent states from the same module.  Of course, we should be wary of the possibility that two modules whose highest weights differ by integers are both present.  However, when the possible fusion products are known, and each has a unique highest weight with none differing by an integer, this uniquely identifies a conformal block with each choice of $\IP$.

\begin{figure}[htbp] 
\centering
\includegraphics*[width=0.3\columnwidth]{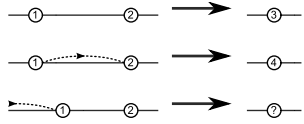} 
\caption{\label{fig:Vrtx_Fusion}
The dashed line represents possible integration paths for $\mathcal{V}_{\pm}$.  Then the top two choices yield distinct $\Phi_1 \times \Phi_2$ fusion channels with charges $\a_3=\a_1+\a_2$ and  $\a_4=\a_1+\a_2+\a_{\pm}$.  The bottom is ambiguous, as the screening charge is not local with respect to the fusion. }

\end{figure}

So we aim to identify fusions where the resulting local charge is unambiguous, and the fusion is to a specific module (see \fref{fig:Vrtx_Fusion}, top two rows).
For pairs of operators that have screening charges entwined with distant points (see \fref{fig:Vrtx_Fusion}, bottom row) there may be multiple fusion products, which need to be identified by other means.

As we mentioned at the beginning of our discussion, the Coulomb gas has a richer structure than our CFT.  This means that while generally we may identify the charge of a fusion product, we can't identify the full structure of the resulting module. In what follows we only identify sets of vertex operators with specific Kac operators for the purposes of generating solutions to \eref{equ:2N_Correlation_function}, and throughout we assume that these identifications are only accurate when the indicated fusion channels are constructed so that they are consistent with the CFT fusion rules.

\subsection{A boundary condition example}

In order to gain an understanding of how a vertex operator representation of  \eref{equ:2N_Correlation_function} with a given choice of integration paths forms a conformal block, we discuss the four--point functions in detail.  This also illustrates a transparent method of calculating the weights of arc configurations via CG.

Since we should expect the expressions for our arc weights to vary continuously as we change $n$, we need our CG expressions to give consistent conformal blocks of \eref{equ:2N_Correlation_function} for arbitrary values of $n$.   The vertex operator configurations for \eref{equ:2N_Correlation_function} that obey charge neutrality for arbitrary $n$ have one $\mathcal{V}_{\op}^+$ vertex, $(2N-1)$ $\mathcal{V}_\op^-$ vertices and $(N-1)$ $Q_-^{\mathcal{I}}$ screening charges.  Keeping to this set of operators will prevent us from getting spurious solutions without meaningful interpretations.

There are only two possible ways to construct the identity and two-leg channel intervals with the permitted vertex operators.  In the following figures we use circles around $\pm$ to represent $\mathcal{V}_{\op}^\pm$ and a dashed line to represent the integration path for any $Q_-$ screening charges.

The simplest arc configuration describes a single arc between boundary points $x_1$ and $x_2$.  This partition function is proportional to the two point function
\begin{equation*} \label{equ:2pnt}
Z_{12}
\sim \langle \f{1} \f{2} \rangle=\langle \vn{1} \vp{2} \rangle\; ,
\end{equation*}
and to the arc weight $W_{12}$ of arcs between $x_1$ and $x_2$. Adding an identity channel boundary condition between $x_1$ and $x_2$ gives an extra loop, thus an extra factor of $n$:
\begin{equation*}
Z_{12}=F_{12}^{12} W_{12}=n W_{12}\; .
\end{equation*}

\begin{figure}[htbp] 
\centering
\makebox[0.15 \columnwidth]{\raisebox{-.2ex}{$\left| 12\lsept34\ket \Rightarrow$}} 
$\Big|\includegraphics*[width=0.15\columnwidth]{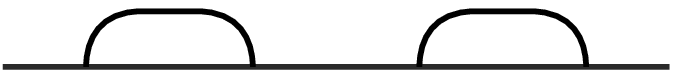}\Big\rangle$
\qquad
\makebox[0.15 \columnwidth]{\raisebox{-.2ex}{$\left| 14\lsept32\ket \Rightarrow$}} 
$\Big|\includegraphics*[width=0.15\columnwidth]{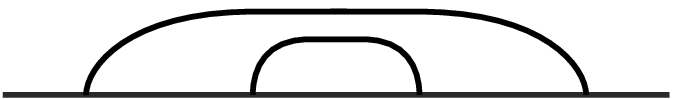}\Big\rangle$
   \caption{ \label{fig:4pt_opa_Instates}
Schematic representation of arc class instates.}
\end{figure}

Though the proportionality between $Z_{12}$ and $W_{12}$ seems somewhat arbitrary, we'll see that it's natural for the four point function. Assuming $x_4<x_3<x_2<x_1$, consider
\begin{equation*}
Z_\IP \nonumber= \langle \f{1} \f{2} \f{3} \f{4} \rangle\sim \langle Q_{-}^\IP \vn{4} \vp{3} \vn{2} \vn{1} \rangle\; ,
\end{equation*}
There are only two arc weights that contribute to these correlation functions. For each arc class $\omega \in \Omega$ we define the usual in-state $\keto{\omega}$ to be the schematic of the bulk arcs in the upper half plane, as in \fref{fig:4pt_opa_Instates}.

\begin{figure}[htbp] 
\centering
\begin{eqnarray*}
\fl
\makebox[0.13 \columnwidth]{\raisebox{.7ex}{$\bra \widehat{12}\lsept\widehat{34}\right|  \Rightarrow$}} 
\includegraphics*[width=0.35\columnwidth]{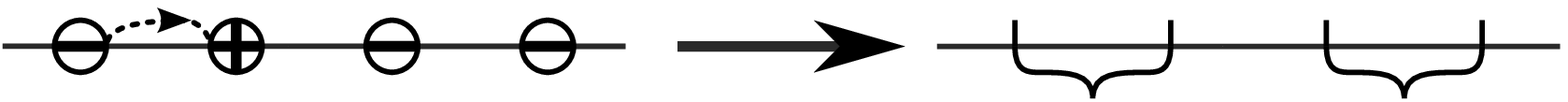}
&
\hspace{3ex}
\makebox[0.13 \columnwidth]{\raisebox{0.5ex}{$\bra 12 \lsept 34
\right| \Rightarrow$}} 
\includegraphics*[width=0.35\columnwidth]{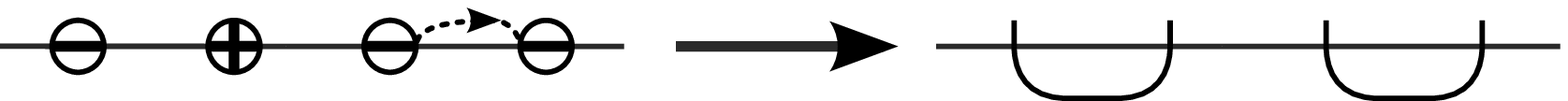}
\\[1ex]
\fl
\makebox[0.13 \columnwidth]{\raisebox{1.2ex}{$\bra  \widehat{14} \lsept \widehat{32} \right| \Rightarrow$}} 
\includegraphics*[width=0.35\columnwidth]{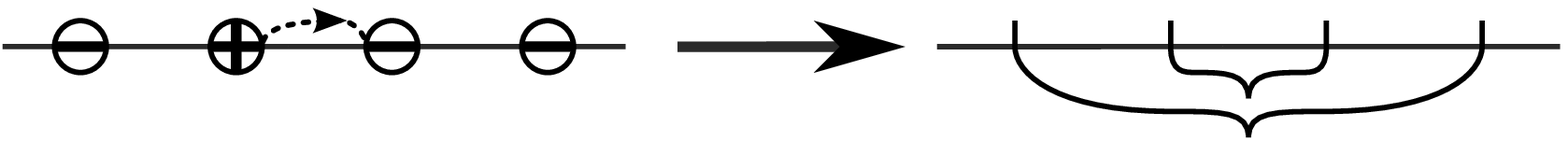}
&
\hspace{3ex}
\makebox[0.13\columnwidth]{\raisebox{0.8ex}{$\bra 14 \lsept 32
\right|  \Rightarrow$}} 
\includegraphics*[width=0.35\columnwidth]{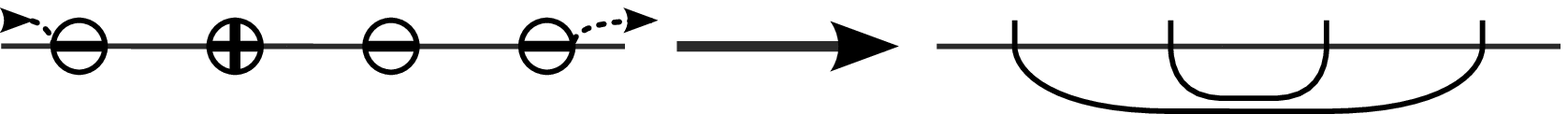}
\end{eqnarray*}
   \caption{ \label{fig:4pt_opa_blocks}
Boundary condition out-states associated with the fusion channels for each conformal block.
}
\end{figure}

\begin{figure}[htbp] 
\centering
\begin{tabular}{rr@{\;}lr@{\;}lr@{\;}lr@{\;}l}
&
\multicolumn{2}{c}{$\bra  \widehat{12}\lsept \widehat{34} \right|$\phantom{\qquad}}&
\multicolumn{2}{c}{$\bra  \widehat{14}\lsept \widehat{32} \right|$\phantom{\qquad}} &
$\left| 14\lsept32\ket$&&
$\left| 14\lsept32\ket$&
\\[1ex]
\raisebox{0.75ex}{$\left| 12\lsept34\ket$\phantom{\quad}}&
\includegraphics*[width=0.12\columnwidth]{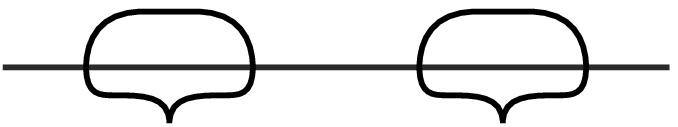}&
\makebox[0.045 \columnwidth]{\raisebox{0.5ex}{$=0$\;}}&
\raisebox{-0.6ex}{\includegraphics*[width=0.12\columnwidth]{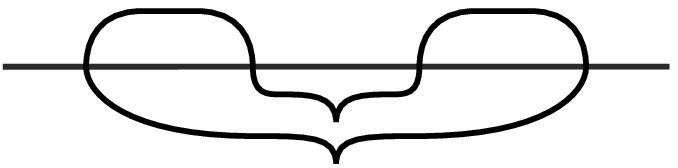}}&
\makebox[0.045 \columnwidth]{\raisebox{0.5ex}{$=n\;$}} &
\includegraphics*[width=0.12\columnwidth]{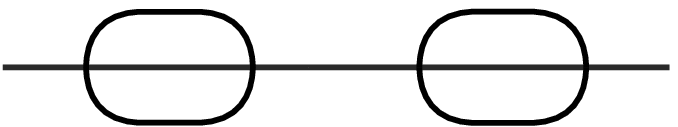}&
\makebox[0.045 \columnwidth]{\raisebox{0.5ex}{$=n^2$\;}}& 
\raisebox{-0.15ex}{\includegraphics*[width=0.12\columnwidth]{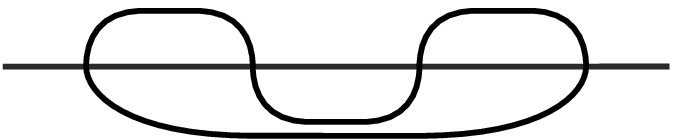}}&
\makebox[0.045 \columnwidth]{\raisebox{0.5ex}{$=n\phantom{^2}$}}
	\\[2ex]
\raisebox{1.25ex}{$\left| 14\lsept32\ket$\phantom{\quad}}& 
\includegraphics*[width=0.12\columnwidth]{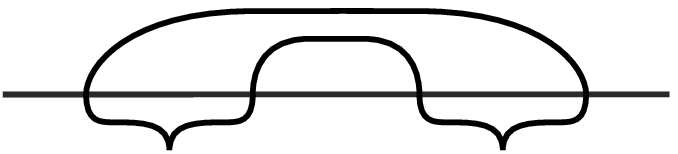}&
\makebox[0.045 \columnwidth]{\raisebox{0.5ex}{$=n$\;}}&
\raisebox{-0.6ex}{\includegraphics*[width=0.12\columnwidth]{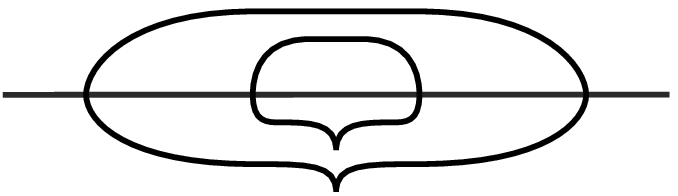}}&
\makebox[0.045 \columnwidth]{\raisebox{0.5ex}{$=0\;$}}&
\includegraphics*[width=0.12\columnwidth]{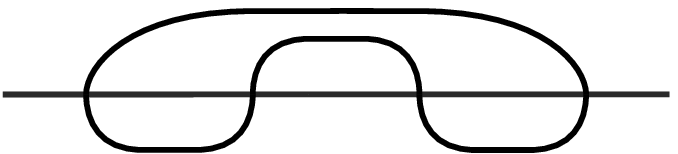}&
\makebox[0.045 \columnwidth]{\raisebox{0.5ex}{$=n\phantom{^2}$\;}}&
\raisebox{-0.15ex}{\includegraphics*[width=0.12\columnwidth]{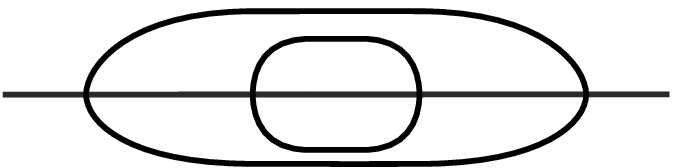}}&
\makebox[0.045 \columnwidth]{\raisebox{0.5ex}{$=n^2$}}
\end{tabular}
   \caption{ \label{fig:4pt_opa_InProd}
The bracket indicates that contractible loops between these two points are forbidden in the corresponding conformal block.
}
\end{figure}

In order to calculate the coefficient of the arc weights in each conformal block we just take the usual meander inner-product \cite{DiFrancesco1998} between each arc in-state and the boundary condition out-states from \fref{fig:4pt_opa_blocks}.  We use the notation $\ldots \lsept ij\lsept \ldots$ to label in- or out-states where $x_i$ is connected to $x_j$ and so on. 
In out-states we place a hat over pairs connected by a bracket in a $\opa$ channel interval. Evaluating the coefficients
\begin{equation*}
F_{BC}^\omega = \bra BC | \omega \ket
\end{equation*}
is a matter of drawing the arcs and boundary conditions together and giving each resulting closed loop a weight $n$ as in \fref{fig:4pt_opa_InProd}.
Assembling these coefficients together with the arc weights gives
\begin{equation} \label{BC_CXing_Relations}
\eqalign{
Z_{\widehat{12} \,\lsept\, \widehat{34}}&=n W_{14 \,\lsept\, 32}\\
Z_{\widehat{14} \,\lsept\, \widehat{32}}&=n W_{12 \,\lsept\, 34}\\
Z_{12 \,\lsept\, 34}&=n^2 W_{12 \,\lsept\, 34}+n W_{14 \,\lsept\, 32}=n  Z_{\widehat{14} \,\lsept\, \widehat{32}}+Z_{\widehat{12} \,\lsept\, \widehat{34}}\\
Z_{14 \,\lsept\, 32}&=n W_{12 \,\lsept\, 34}+n^2 W_{14 \,\lsept\, 32}=Z_{\widehat{14} \,\lsept\, \widehat{32}}+n Z_{\widehat{12} \,\lsept\, \widehat{34}}
}
\end{equation}

Dotsenko and Fateev calculated crossing relations between these four conformal blocks in \cite{Dotsenko1984}.
Integrating the screening charge over a closed loop in the upper half plane the result would be zero.  We deform this loop onto the real axis so that it passes slightly above the vertex operators.  Now the integral, which is still zero, is a sum of the real conformal blocks from \fref{fig:4pt_opa_blocks} with a phase picked up each time the integration proceeds over an operator.  Passing over  ${\cal V}_{1,2}^-$ (resp.~${\cal V}_{1,2}^+$) yields a phase of $4 \pi/\k$ (resp.~$-12 \pi /\k$).  If we start with a phase $\chi$ we get the result
\begin{equation} \label{equ:4pt_CB_Relation}
0=\E^{\I \chi}\left(
Z_{14 \lsep 32}
+\E^{\I 4 \pi/\k}Z_{\widehat{12}\, \lsep\, \widehat{34}}
+\E^{-\I 8 \pi/\k}Z_{\widehat{14}\, \lsep \,\widehat{32}}
+\E^{-\I 4 \pi/\k}Z_{12 \lsep 34} \right)\; .
\end{equation}
For $\chi=0$ or $4 \pi/ \k$, taking the imaginary part of the expression and simplifying with (\ref{n_of_kappa}) gives the expressions in \eref{BC_CXing_Relations}. As we alluded to in \fref{fig:CvR_Contour} the real integral paths only work when $\k>4$ and the integral along the real line converges.   However, it can be shown with just a little more work that the results extend to $0<\k \le4$ as well.

For the four--point function, there are only two arc weights that contribute to four conformal blocks.   This correlation function is the simplest that exhibits both fusion channels $\op \times \op = \M_1\oplus \M_{\opa}$.  For the percolation model $\opa$ is the pre-logarithmic operator that precedes  the staggered modules we study and the four point function can be used as the starting point for calculating arc weights for a greater number of arcs.  In the next section we'll describe how to add identity channel intervals to this CG expression to make such a calculation.

\subsection{Identity channel intervals}

Take a properly normalized CG representation of the $2N$--point correlation function 
\begin{equation*}
\bra \prod_{i=1}^{2N} \op(x_i) \ket
=C \bra\vp{1} \prod_{j=2}^{2 N} \vn{j} \prod_{k=1}^{N-1} Q_-^{\IP_k} \ket,
\end{equation*}
where our choice of $\IP_k$ fixes the conformal block uniquely by the conventions in the previous section.
For simplicity we denote the collection of vertices and screening charges by $\mathcal{O}$.  To this correlation function we add the local charge--neutral object 
\be \label{equ: Id_Channel_CG_Ops}
\K Q_{-}^{(x_a,x_b)} \vn{a}\vn{b}\; ,
\ee
with $x_a$ and $x_b$ adjacent on the real axis and $\K$ a fixed constant. The fact that \eref{equ: Id_Channel_CG_Ops} is charge neutral means that when $x_a \to x_b$ we recover the original $2N$--point function with the fusion channels between the remaining operators unchanged.  Inserting this object is equivalent to adding an $\{\op\op\}_1$ to the original correlation function, without changing the existing fusion channels.

We can identify the proper normalization of the resulting $2(N+1)$--point function \cite{Dubedat2006}.  The leading fusion product as we contract the $\{\op\op\}_1$ is the identity, so when $x_a-x_b \ll x_a-x_j$ for all $1 \le j \le2N$ we can write
\begin{equation*}
\fl
\bra \left\{ \op(x_a)\op(x_b)\right\}_1 \prod \op(x_j)\ket\nn
=\bra \op(x_a)\op(x_b) \ket
\left[
\bra \prod \op(x_j) \ket 
+ \mathrm{O}(x_a-x_b)\right]\; ,
\end{equation*}
or in terms of the vertex operators
\begin{equation*}
\bra \K Q_{-}^{(x_a,x_b)} \vn{a}\vn{b}
\mathcal{O}
\ket
=
\left(x_b-x_a\right)^{-2h_{1,2}}
\left[
\bra \mathcal{O} \ket + \mathrm{O}(x_a-x_b)\right]\; .
\end{equation*}
Expanding the integral expressions to first order in small parameters yields
\begin{equation*}
\K\int_{x_a}^{x_b} \frac{(x_b-x_a)^{2 \left(\a_{1,2}^-\right)^2} \mathrm{d}u}{[(x_b-u)(u-x_a)]^{-2\a^- \a_{1,2}^-}} \bra \mathcal{O} \ket 
=\nn
(x_b-x_a)^{-2h_{1,2}}\bra\mathcal{O}\ket \; ,
\end{equation*}
which allows us to fix $\K$:
\begin{equation} \label{equ:K_derivation}
\fl
\K
=(x_b-x_a)^{1-6/\k}\left[ \int_{x_a}^{x_b} \frac{(x_b-x_a)^{2/\k} \mathrm{d}u}{[(x_b-u)(u-x_a)]^{4/\k}} \right]^{-1}
=\frac{\Gamma(2-8/\k)}{\Gamma(1-4/\k)^2}
\; .
\end{equation}
This lets us add an \ICI~to a given $2N$--point block and immediately write down a CG expression for a new $2(1+N)$--point block.

Each time we add a new \ICI~we extend the set of allowed arc classes from $\Omega^a_N \Rightarrow \Omega^a_{N+1}$.  An $(N+1)$ arc class belongs to $\Omega_{N+1}^a$ if either, the ends of the new \ICI~are connected by a contractible arc and erasing this arc gives an element in $\Omega_N^a$, or if the new \ICI~is connected to propagating arcs and identifying the ends of these arcs gives an element in $\Omega_N^a$.  These cases are both simple results of the identity channel interval boundary condition. In \sref{sec: 3 Interval Xing} we'll discuss the three arc case in detail.

\section{Percolation \label{sec:Perc}}

In the remainder of this paper we focus on the percolation model and the $c=0$ LCFT that describes it.  We'll begin with a brief review of the percolation model, followed by an equally brief description of pertinent LCFT operators.  Then we'll describe a variety of boundary intervals that are both simple to construct in the percolation model and can be identified with LCFT fusion channels.

\begin{figure}[htbp] 
\centering
\hspace*{\stretch{1}}
\includegraphics*[width=0.28\columnwidth]{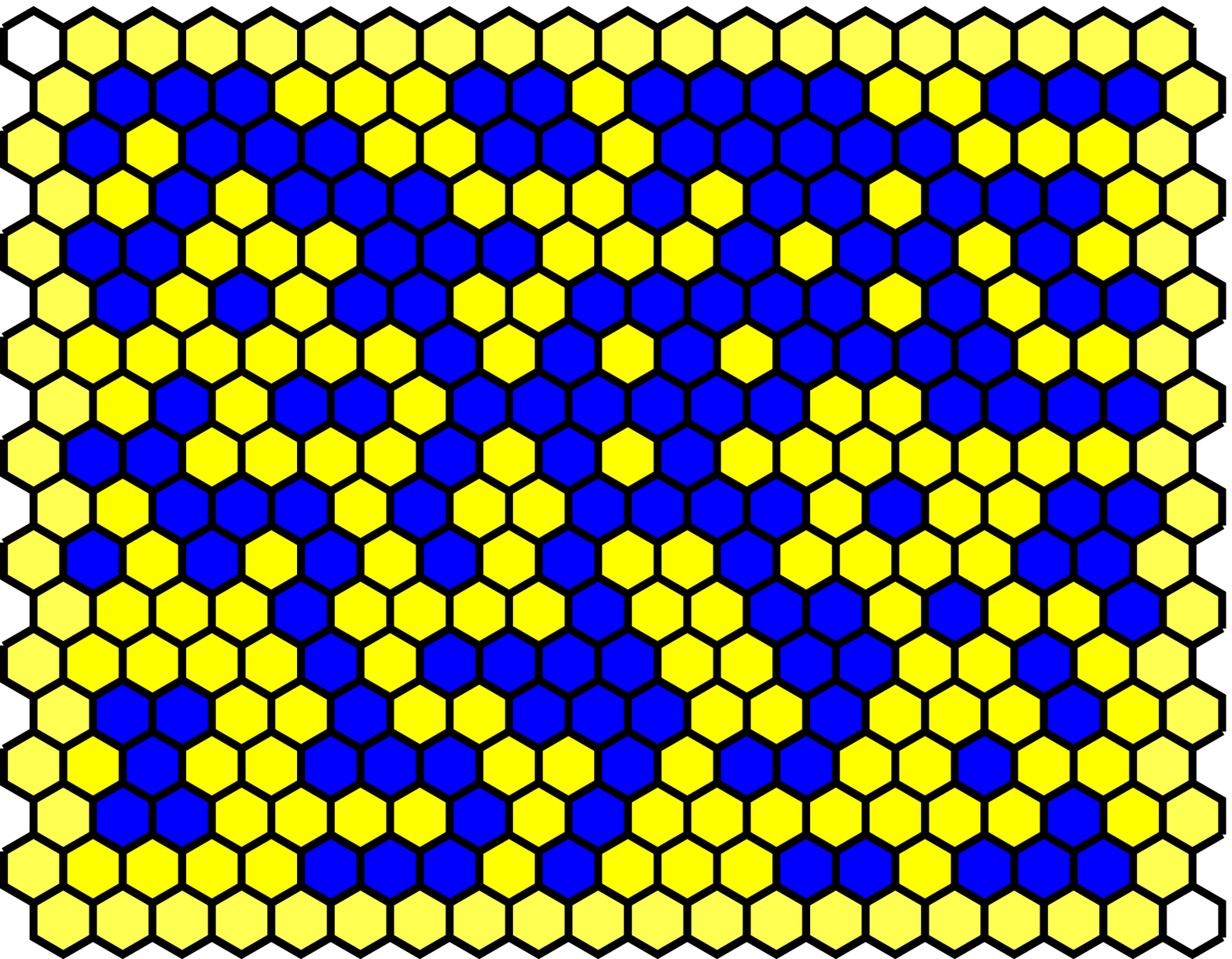}
\hspace*{\stretch{1}} 
\includegraphics*[width=0.28\columnwidth]{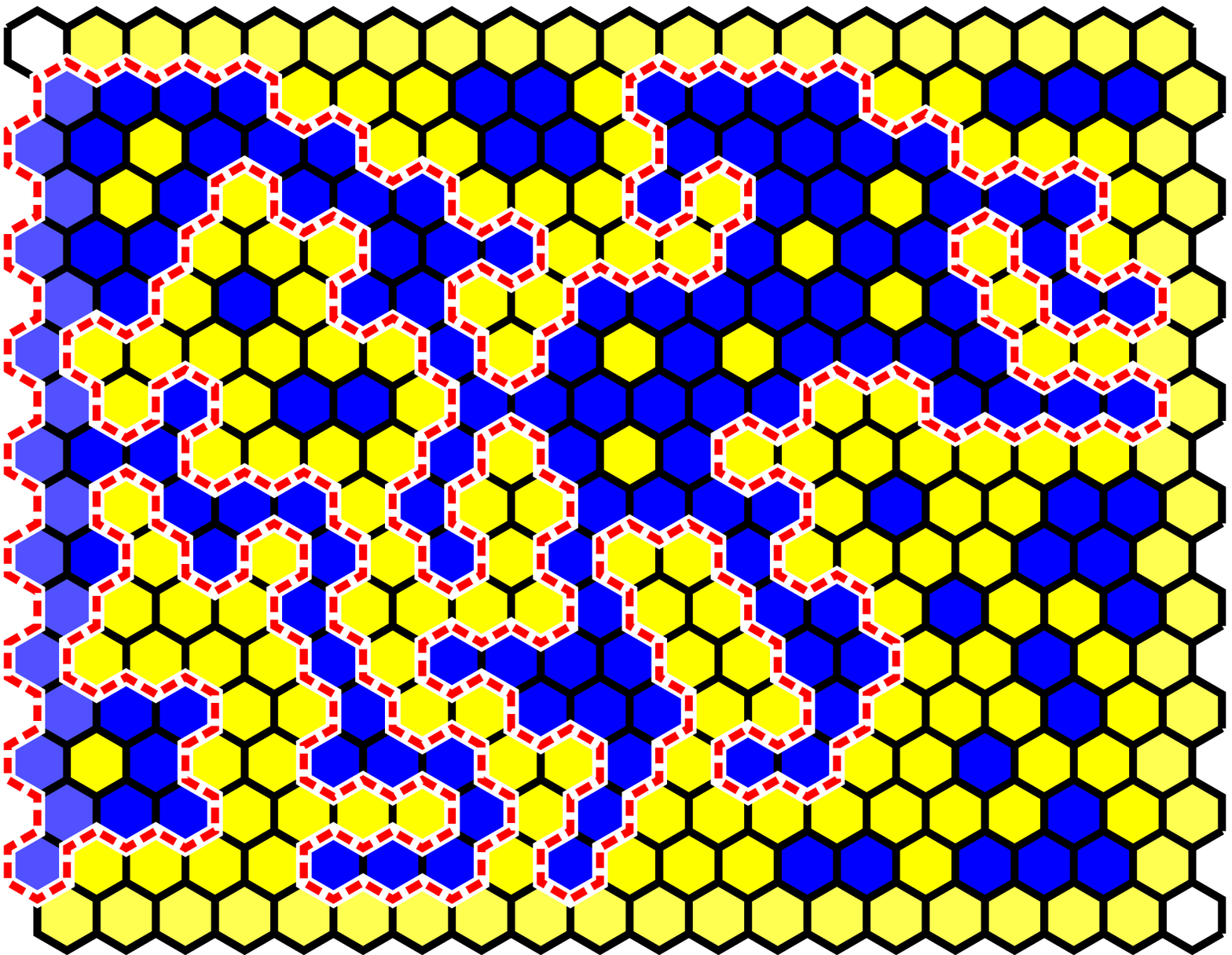}
\hspace*{\stretch{1}}
\caption{\label{fig:Perc_v_SLE}
LEFT: a schematic of an ${\rm O}(n)$ configuration with homogeneous boundary conditions. RIGHT: an illustration of a configuration with an SLE loop, which extends to enclose the fixed boundary segment.
}
\end{figure}

Perhaps the easiest formulation of the model is formed by taking the sites of a triangular lattice and randomly coloring each site blue or yellow with equal probability to get configurations like those in figure \ref{fig:Perc_v_SLE}.
  In terms of the $O(n)$ loop model this correspond to the $n=1$ dense phase, and the ``loops'' are the domain walls between between yellow and blue regions on the graph.

We call the blue sites \emph{occupied} and groups of adjacent blue sites \emph{clusters}.    In this convention we'd color the boundary yellow to represent a \emph{free} boundary (as shown in \fref{fig:Perc_v_SLE}). Clusters that abut a free boundary maintain their unique identity.  Alternately, we color a \emph{fixed} boundary blue.  All the clusters touching a fixed boundary are considered part of a single boundary cluster.  Of course, choosing blue clusters was arbitrary and the role of yellow and blue can be reversed.

In the CFT for percolation, the boundary operators $\op(x_a)$ mark changes from yellow to blue boundary sites.  We can uniquely identify the path along the outer edge of the clusters attached to the blue boundary for each realization of the bulk state (see the right side of figure \ref{fig:Perc_v_SLE} for an example of a single pair of boundary operators) and this objects forms the arcs we discussed earlier.  If we include multiple boundary operators, then for each added interval we'll see another arc in each bulk configuration.

However, the state of each bulk site is i.i.d., so the weights of the bulk configurations are independent of our choice of boundary conditions.  This is novel among the statistical mechanics models that can be mapped to the $O(n)$ loop model.  In the SLE literature it's a well know result called \emph{locality} \cite{LawlerSchrammWerner01} that holds only for $\k=6$ and states that the growth of $SLE_6$ traces do not depend on the boundary unless they collide with it.

The implication for the percolation model is that the bulk weights are entirely unaffected by our choice of boundary conditions, which only serve to decompose configurations into arc classes.  This is in contrast to the $O(n)$ model where changes in the boundary conditions change both the structure of the arc classes and the weight assigned to each configuration.   The ability to compare different sets of boundary operators leads to a straightforward correlation between arc configurations and the operators in the $c=0$ LCFT.

\subsection{A brief review of necessary $c=0$ LCFT operator}

The minimal $c=0$ CFT that describes the local percolation theory is trivial, which is expected from the locality property. Only two operators appear: the identity and the blue/yellow boundary change operator $\op$.  Both have weight zero and are annihilated by Virasoro generators $L_{-1}$ and $\left(2L_{-2}-3L_{-1}^2\right)$.  Using the commutation relations for the Virasoro generators, it's easy to show that all the descendants of a primary operator are zero if they're annihilated by  both these generators.

However the class of observables we consider is non-local and depends on the arc weights of the domain wall theory instead.   In the non-local theory the identity is still annihilated by $L_{-1}$ but not the second order generator.  This means that while the state $L_{-2}\keto{1}=\keto{T}$ is still singular it is not null.  Non-null singular states have logarithmic partners, in this case the partner to the stress tensor is $\phi_{1,5}=:\Tlog$ \cite{MathieuRidout07}.

These operators belong to the staggered module $\Imod_{1,5}=:\Imod_{\Tlog}$ with cyclic operator $\Tlog$ and highest weight operator 1. Significant relations in this staggered module include
\begin{equation*}
\begin{array}{l@{\,}ll@{\,}ll@{\,}ll@{\,}l}
L_0\keto{1}&=0&
 L_0\keto{T}&=2\keto{T}&
L_0 \keto{\Tlog}&=2\keto{\Tlog}+\keto{T}&
L_2 \keto{\Tlog}&=-\frac58 \keto{1}\; .
\end{array}
\end{equation*}

Meanwhile, SLE tells us that the second order generator must still annihilate $\keto{\op}$ \cite{BauerBernard2003}, but $L_{-1}\keto{\op}=\pop\neq0$ if we're to have non-trivial results.  So $\pop$ is singular but not null and couples to a logarithmic partner $\phi_{1,4}=:\poplog$.  These operators belong to the staggered module $\Imod_{1,4}=:\Imod_{\poplog}$ with cyclic operator $\poplog$ and highest weight operator $\op$.  Significant relations in this staggered module include
\begin{equation*}
\begin{array}{l@{\,}ll@{\,}ll@{\,}ll@{\,}l}
L_0 \keto{\op}&=0&
L_0 \keto{\pop}&=\keto{\pop}&
L_0 \keto{\poplog}&=\keto{\poplog}+\keto{\pop}&
L_1 \keto{\poplog}&=-\frac12\keto{\op} \; .
\end{array}
\end{equation*}

At this level considered in this paper, the final important operator is the two-leg operator $\phi_{1,3}=\opa$ that was discussed in the context of the $O(n)$ model.  In the $c=0$ model this operator belongs to the edge of the Kac table, so that $\M_\opasub=\Lmod_\opasub$. Furthermore, the operator is pre-logarithmic; while $\opa$ is a regular operator, its appearance leads to the logarithmic modules in the fusion rules of the theory.

The operators we consider are all found within the first three fusions of the boundary operator.  The relevant fusion rules are
\begin{equation*}
\op{}^2 = \M_{1}\oplus\Lmod_{\opa},\quad
\op{}^3 = \M_{\op}\oplus\Imod_{\poplog}, \quad \textrm{ and}\quad
\op{}^4 = \M_{1}\oplus3\, \Lmod_{\opa}\oplus\Imod_{\tau},\;.
\end{equation*}
We will not comment on how the sub-sector generated by $\op$ fits into a more complete theory.  In particular we won't address the interaction with the bulk sector.

\subsection{Identity v. two-leg intervals}

In the remainder of this section we discuss fusion channel intervals: intervals with specific boundary operator constructions that lead to unique fusion channels when contracted to a single point.  We'll construct intervals for the five operators discussed above ($\opa, \pop, \poplog, T,$ and $\Tlog$), and describe how they apply to percolation crossing events.   We pick to work with intervals because this is perhaps the easiest way to imagine all the significant operators as objects on equal footing, and because it makes constructing higher order correlation functions through the Coulomb gas relatively simple.

\begin{figure}[ht] 
\centering
\makebox[0.1 \columnwidth]{\raisebox{2ex}{$\{ \op \op \}_{\idsub}\; :$}\hspace{\stretch{1}}}
 \includegraphics*[width=0.15\columnwidth]{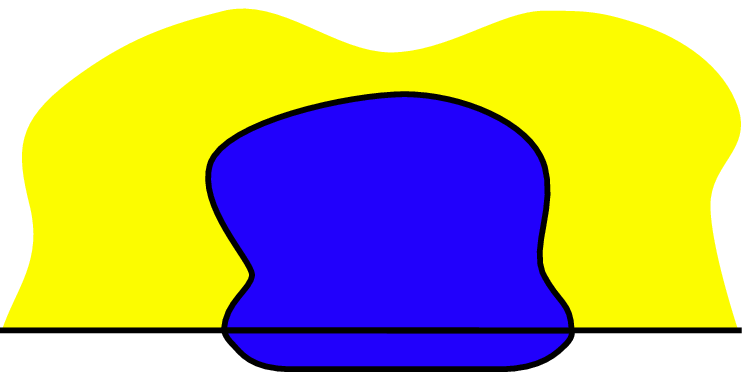}
\raisebox{2ex}{\huge\;$\cup$\;}
\includegraphics*[width=0.15\columnwidth]{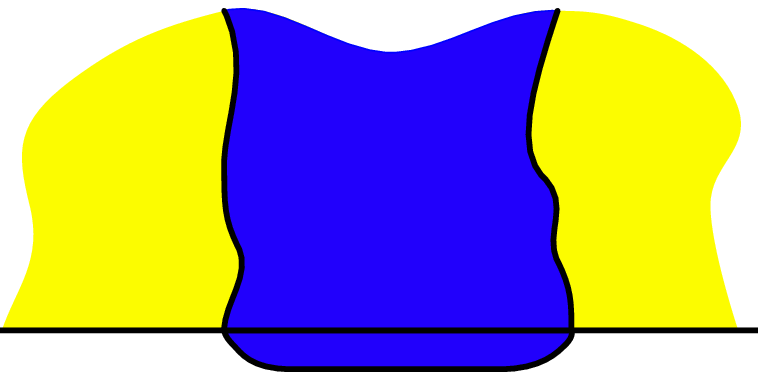}
\qquad
\makebox[0.1 \columnwidth]{\raisebox{2ex}{$\{ \op \op \}_{\opa}\; :$}\hspace{\stretch{1}}}
\includegraphics*[width=0.16\columnwidth]{Figure9b}
\caption{ \label{fig:Perc_IdInt}
Contractible and propagating arcs both contribute to the identity interval, while only propagating arcs contribute to the two--leg interval.
}
\end{figure}

The identity and two-leg intervals from section \ref{sec: O n Loop} are easily specialized to the percolation model.  They can both be formed by coloring a portion of the boundary blue (or yellow if need be).  The arc defined by these boundary sites is either contractible or propagating with respect to the interval as before.  The identity channel allows both contractible and propagating arcs, while the two-leg interval only allows arcs that propagate toward another two-leg interval, both cases are illustrated schematically in \fref {fig:Perc_IdInt}.

In percolation, the meaning of a contractible arc is simple: no cluster touching the interval connects to any other boundary change intervals.   Propagating arcs are more complicated because fixing the interval changes the connectivity of incident clusters and when calculating crossing probabilities, one treats connections made through bulk clusters differently from connections that rely on fixed boundary intervals \cite{SKFZ2011}.  These incidental connections, are at the heart of the LCFT operator correspondence.

\subsection{Two-point fusion: the stress-tensor interval}

As is usual in CFT, the identity module $\M_1$ is comprised of the identity operator, and the stress tensor $T$ with its descendants.  In the $c=0$ theory the stress tensor is also a singular primary operator, which generates the maximal submodule of $\M_\idsub$. If we quotient the identity operator from $\M_\idsub$, we're left with this irreducible submodule, which we call $\Lmod_T:=\M_1\backslash 1$.

The fusion rules of the theory do not produce this submodule in isolation.  However, it is easy to isolate its contribution within an \ICI~interval.  For any correlation function containing an \ICI~pair, we just subtract the correlation function where this pair is absent.  Of course, we must be careful to leave all other operators and fusion channels unchanged. We denote this difference as $\{\op \op\}_T:=\{\op\op\}_1-1$, and call it a \emph{stress tensor channel interval}.

If we start with a correlation function that gives the weight of some observable, then adding a stress tensor interval is equivalent to asking how the weight responds to changing the boundary conditions on the interval. Lets say the original observable can be calculated from an $2N$--point function like \eref{equ:2N_Correlation_function}, and that occurrences of the observable mean that the arc class of the configuration belongs to a particular subset of the $N$--arc classes: $[\Lambda]\in\Omega_N^a\subset\Omega_N$, where the bracket $[\bullet]_N$ denotes the equivalence class in $\Omega_N$.  Then, as we discussed, adding \ICI~naturally extends the subset $\Omega_N^a\Rightarrow \Omega_{N+1}^a\subset\Omega_{N+1}$.  The version of the correlation function that includes the \STI~equals the weight of configurations $\Lambda$ such that $[\Lambda]_{N+1}\in \Omega_{N+1}^a$ and $[\Lambda]_{N} \notin \Omega_{N}^a$ minus the weight of configurations $M$ such that $[M]_{N+1}\notin \Omega_{N+1}^a$ and $[M]_{N} \in \Omega_{N}^a$.

Now cluster observable in critical percolation require that either a crossing cluster connects a set of regions, or that the clusters attached to these regions be disjoint. This means that the change of boundary conditions associated to the \STI~is only important in configurations where the clusters attached to these regions are disjoint in the bulk but happen to both be incident on the location of the \STI~interval.  The the observable will manifest disjoint clusters if the boundary sites are yellow, and crossing clusters when the boundary sites are blue.

\begin{figure}[ht] 
\centering
\makebox[0.1 \columnwidth]{\raisebox{2ex}{$\{ \op \op \}_{T}\; :$}\hspace{\stretch{1}}}
 \includegraphics*[width=0.17\columnwidth]{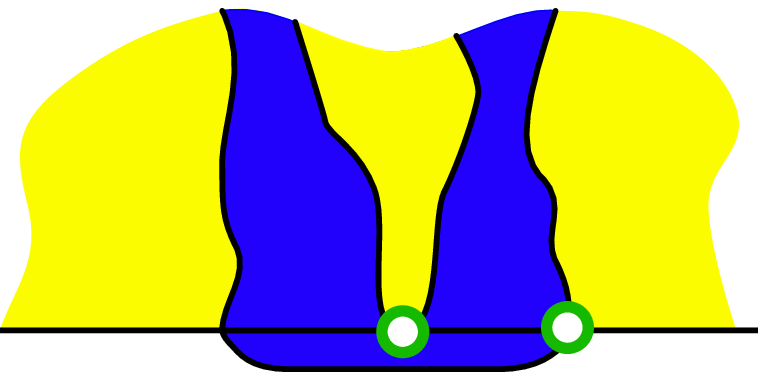}
\raisebox{2ex}{\huge\;$-$\;}
\includegraphics*[width=0.17\columnwidth]{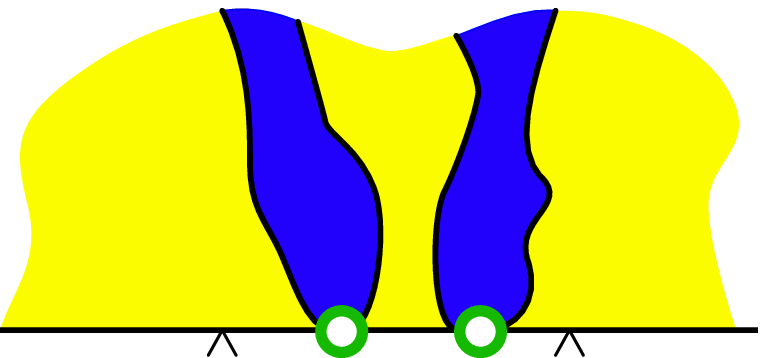}
   \caption{ \label{fig:Perc_TInt}
If two clusters are important to the original observable $\langle \mathcal{O} \rangle$, disjoint in the bulk, and both touch the new \STI~interval, then the new correlation function $\langle \{\op\op\}_T \mathcal{O}\rangle$ receives a positive (resp.~negative) contribution if the clusters are required to connect (resp.~be disjoint) in $\langle \mathcal{O} \rangle$.  Of course, all the properties of arcs that are not incident on the new interval (and not included in this figure) remain the same.
}
\end{figure}

The configurations that contribute to the stress tensor channel interval are illustrated in figure \ref{fig:Perc_TInt}, for the case where we add a blue \ICI~interval to a yellow boundary.
In the figure, the `$\circ$' indicates that the portions of the arcs shown belong to distinct arcs from the original observable and that the schematic representations of the arc classes are identical before and after the \STI~insertion, with the obvious exception that some pairs of arcs are conditioned to surround or touch the new boundary interval.  Because the target arc configurations are unchanged by the \STI~insertion, we refer to it as \emph{passive} insertion.

It's important to note that only configurations with at least two distinct arcs from the original correlation function touching the added \STI~can contribute.  If a single arc touches the interval then while the shape of that arc changes, the arc class of the correlation function will stay the same.  The two arc requirement is reflective of the CFT; $T$ only couples to it's logarithmic partner $\tau$, and it takes a minimum of four $\op$ operators (which generate two arcs) to produce the $\Imod_\tau$ module.

In the Coulomb gas it's easy to input the stress tensor interval.  Recall the one can add an identity channel to a conformal block without changing the existing fusion channels by adding the charge neutral object (\ref{equ: Id_Channel_CG_Ops}).  The an insertion of \STI~is found by taking the difference between the modified and original CG expressions.

\begin{figure}[ht] 
\centering
\includegraphics*[width=0.2\columnwidth]{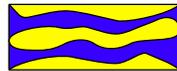}
\caption{ \label{fig:multi_4-pnt}
These multiple crossings superficially resemble two \STI 's.  However, the loop surrounding the internal yellow region is inconsistent with this interpretation; its inclusion requires elements from $\Imod_\Tlogsub$.
}
\end{figure}

Superficially, it may seem that two \STI~intervals could couple to each other as in figure \ref{fig:multi_4-pnt}, 
but $T$ is a singular operator and does not couple to itself, so it's two point function is zero. The resolution to this apparent inconsistency comes from the fact that \STI~is a passive operator.  Since a \STI~only interacts with existing arcs in the observable to which it's added, any correlation function made up entirely of \STI, must be zero, since effectively there will be no arcs in the associated arc class.  In \sref{sec: STPI} we'll discuss how to build and interpret correlation functions associated to the configurations seen in \fref{fig:multi_4-pnt}, in particular see \fref{fig:Expected No-1_Convention}.

\subsection{The derivative of the boundary change operator \label{pop}}

The straight forward way to construct an interval $(x_0, x_1)$ that fuses to the $\pop$ is by taking the difference of two correlation functions
\be
\bra \op(x_1) \mathcal{O}\ket-\bra \op(x_0) \mathcal{O}\ket\; , \qquad \textrm{where}\quad \mathcal{O}=\prod_{i=2}^{2n} \op(x_i)\;,
\ee
and we isolate the same conformal blocks in both correlation functions.We use the notation $\Delta\op(x_a,x_b):=\op(x_b)-\op(x_a)$ as shorthand for this interval.

This is an almost trivial observation, but the corresponding arc classes are not.  As with the stress tensor interval, configurations contribute if their arc class only belongs to the observable subset when $\op$ is at $x_1$ or $x_0$ but not both.

As a result of the locality property, these configurations must have at least one arc from the observable (besides the one attached to the $\op$ that we move) that touches the interval $(x_0, x_1)$.  If this incident arc is required to either connect to or be disjoint from the fixed boundary adjacent to the $\op$ we're moving, then the configuration only contributes to the observable in one of the two positions.  If it contributes to the observable with the bcc in its new position, then it adds a positive weight to the correlation function. If it contributes with the bcc at the original position is contributes a negative weight.

\begin{figure}[ht] 
\centering
\makebox[0.17 \columnwidth]{\raisebox{2ex}{$\Delta\op(x_1,x_2)\; :$}} 
\includegraphics*[width=0.17\columnwidth]{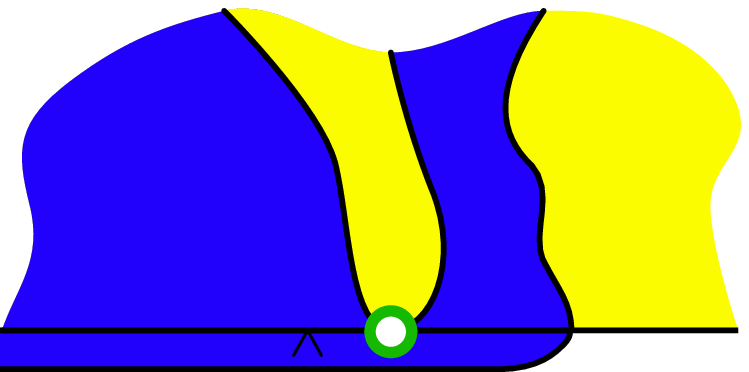}
\raisebox{2ex}{\huge\;--\;}\includegraphics*[width=0.17\columnwidth]{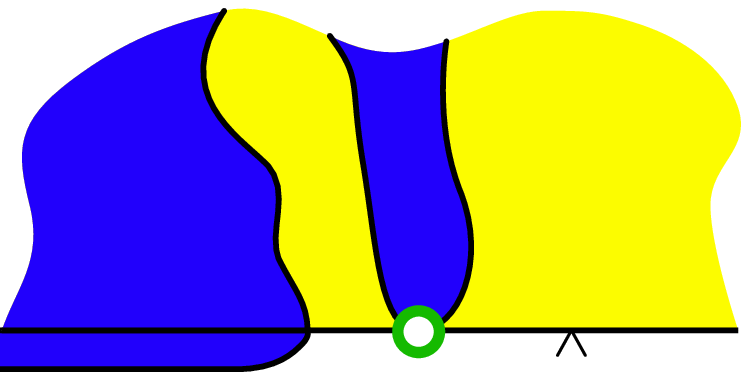}
 \caption{ \label{fig:perc_pop_Convention}
When it's important how a cluster interacts with the boundary interval adjacent to $\op$ in the observable $\langle \op \mathcal{O} \rangle$, and the cluster hits between the two end points of a proposed $\Delta \op$ interval based on $\op$, then the new correlation function $\langle \Delta \op \mathcal{O}\rangle$ receives a positive (resp.~negative) contribution if the clusters are required to hit (resp.~be disjoint from) the fixed boundary in $\langle \mathcal{O} \rangle$.  We assume that all arc properties remain the same before and after the move.}
\end{figure}

A schematic of the $\Delta\op$ interval is shown in \fref{fig:perc_pop_Convention}, in this case the fixed and free boundaries are to the left and right of the $\op$ respectively.  In the figure we assume that the final position is to the right of the initial position.
The `$\circ$' mark emphasizes that this incident arc must satisfy the requirements of the original observable and be a distinct arc from the one that ends at the $\op$ we're moving.  The caret in the figure marks the alternate position of the $\op$.  For an example of this construction consider the discussion surrounding \eref{equ: DYDf_CorFunc}.

The structure and interpretation of the $\Delta\op$ interval is very similar to the stress tensor interval, except that it is attached to a bcc operator instead of an interval from a homogeneous boundary segment.  Our \STI~discussion of `two regions' whose clusters are either connected or disjoint still applies, except one of the two regions will always be the boundary to one side of the bcc operator.  It's particularly important to notice that $\Delta \op$ is also a passive interval because moving $\op$ won't change the target arc classes, it just requires certain arcs to be incident on the boundary.

It's worth noting that the $\Delta \op$ interval only attaches to observables with an unattached 1--leg operator and at least one additional arc.  It takes three bcc operators to produce a boundary change and an extra boundary arc.  This is significant because $\pop$ only couples to its logarithmic partner $\poplog$, and it takes a minimum of three $\op$ to generate $\Imod_\poplog$.

An issue comes up when we use CG to write correlation functions for percolation.  In this case $g=2/3$, and for this value of $g$ the vertex operator $\mathcal{V}_\op^+$  has a charge $\a_{1,2}^+=0$, and absolutely must act as an identity operator.  In particular we need to be careful when we try and entwine $\mathcal{V}_\op^+$ with a screening charge.  With no branch cut associated to the vertex we can't take the normal integration paths (see \fref{fig:CvR_Contour}) as such paths become `unpinned', form a close loop and lead to trivial correlation functions.

This peculiarity arises because the minimal CFT for $c=0$ is trivial.  In particular, if we can't entwine the $\mathcal{V}_\op^+$ operator with a screening charge then we can't write $\{\op\op\}_\opasub$ intervals that are charge neutral and fusion rule consistent.  So, in some sense the unpinning of these integration paths corresponds to the truncation of $\opa$ from the minimal theory.

However, there is a CG work around for this unpinning issue for the non-minimal theory: use an open integration path along the real line for any screening charges that entwine the $\mathcal{V}_\op^+$.  Effectively, the trivial $\mathcal{V}_\op^+$ is reduced to the role of place holder for the integration.
We then represent the two--leg fusion channel by $\vn{1} Q_{-}^{(x_1,x_2)} \sim \{\op(x_1)\op(x_2)\}_\opa$.

If we take the derivative with respect to $x_2$ note the screening charge becomes a valid local operator in it's own right $\mathcal{V}_{-}(x) \sim \pop(x)$, just so long as there are $(2N+3)$ of the $\mathcal{V}_\op^-$ operators and $N$ of the $Q_-$ screening charges as well.  This interpretation is not surprising since $\pop$ is a primary operator and $\a_-=-2/\sqrt{6}=\a_{1,4}^+$.  We'll denote this vertex operator $\mathcal{V}_{\pop}^+:=\mathcal{V}_{1,4}^+$.

\subsection{Three-point fusion: Regular operator \label{sec: 3ptFuse_Reg}}

There is another significant and distinct way to generate an interval that fuses into a $\pop$ operator.   It requires us to examine the three operator fusion $\op{}^3$ in some detail.  In what follows we present the basic results, where relevant, while reserving the full analysis for \ref{Appendix}.

The module $\M_{\op}$ is comprised of the operator $\op$ and its descendants.   In light of the null state and the general commutation relation $[L_{-1},L_{-n}]=(n-1)L_{-n-1}$, any descendant of $\op$ is in the module generated by $\pop$.  But $\pop$ is also a singular primary operator, so if we quotient $\M_\op$ by $\op$ we get the irreducible module generated by $\pop$ which denote as $\Lmod_{\pop}:=\M_\op\backslash \op$.

When we consider the fusion of three $\op$ boundary operators into the $\M_\op$ module, we get two possibilities depending on whether internally it's the left- or rightmost pair operators that fuse to $\M_1$.  We work through the general form of this operator product in the appendix, but the two routes to the $\M_{\op}$ module are:
\begin{eqnarray}
\left\{\op(\ve)\op(\l\ve)\right\}_1\keto{\op}&=\keto{\op}+\ve\, \big(1-\F^{1-\l}_1 \big) \keto{\pop}+\mathrm{O}(\ve^2)\\
\op(\ve) \left\{\op(\l \ve) \keto{\op}\right\}_1&=\keto{\op}+\ve\,\F^\l_1 \keto{\pop}+\mathrm{O}(\ve^2)\; ,
\end{eqnarray}
where
\begin{equation*}
\F^\l_1=1-\frac{1}{5}\l^2{}_2F_1\left(1,4/3;8/3;\l \right)\;.
\end{equation*}

To leading order these two fusion schemes look the same and taking their difference leaves only contributions from the $\Lmod_{\pop}$ module. 
We'll adopt the bracket notation $\{\op\op\op\}_{\pop}$ for this difference:
\begin{eqnarray} \nonumber
\left\{ \op(\ve) \op(\l \ve) \keto{\op} \right\}_{\pop}&:=
\op(\ve) \left\{\op(\l \ve) \keto{\op}\right\}_1-\left\{\op(\ve)\op(\l\ve)\right\}_1\keto{\op}\\
&\;=2\ve\,\F^\l_3 \keto{\pop}+\mathrm{O}(\ve^2)\; ,
\end{eqnarray}
where
\begin{equation*}
\F^\l_3=\frac{\Gamma(2/3)^2}{\Gamma(1/3)}  
 \left[\l(1-\l)\right]^{1/3}\;.
\end{equation*}

\begin{figure}[ht] 
\centering
\makebox[0.2\columnwidth]{\raisebox{2ex}{$\{ \op \op\op \}_{\pop}\; :$}} 
\includegraphics*[width=0.4\columnwidth]{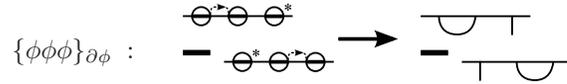}
\caption{ \label{fig:pop_Convention2}
The three point fusion that yields the $\pop$ channel.  Note the similarity between the boundary conditions and the incident arcs in figure \ref{fig:perc_pop_Convention}.}
\end{figure}

The boundary condition associated to this fusion channel is included in \fref{fig:pop_Convention2}.
This boundary condition produces arc classes identical to the $\Delta \op$ interval, except that now they have more internal structure because we've fixed the intermediate anchoring point instead of letting the cluster hit anywhere along the interval.

We can verify that these boundary conditions only couple to $\Imod_\poplogsub$ by verifying that these boundary conditions only couple to three propagating arcs.  Consider a contractible arc connecting the left and middle $\op$'s in \fref{fig:pop_Convention2}, which necessarily leaves a propagating arc attached to the right $\op$.  For the upper boundary condition we get a closed loop from the contractible arc, and the propagating arc connects directly to our dangling boundary operator.  For the lower boundary condition the propagating arc connects to the dangling boundary operator indirectly first via the boundary conditions then via the contractible loop.  But, the weight of the single loop is one, the propagating arc connects to the dangling boundary condition in both cases.  So none of the arc classes change and the net contribution from the difference is zero.  Symmetry gives the same result if the contractible loop attaches the middle and right operators, so these boundary conditions only couples to configurations with non-contractible loops.

We can remove evidence of the three-point $\pop$ interval internal structure if we'd like. Take a small interval of width $\Delta x$ and attach three operators in the $\pop$ fusion channel, as in \fref{fig:pop_Convention2}, then divide by $2 \F^\l_3$.  This leaves a leading contribution of $\Delta x\, \pop$, and if we then stack a large number of these intervals sides by side, say spanning some interval $(0,x)$, and take $\ve \to 0$ we get the Riemann sum
\begin{equation*}
\Delta\op(x_1,x_2):=\op(x_2)-\op(x_1)=\int_{x_1}^{x_2} \pop(w) dw\; .
\end{equation*}
While this seems like a trivial construction it is important; it shows us how to take the logarithmic operator and construct a meaningful logarithmic interval.

\subsection{Three-point fusion: logarithmic operator}

The $\pop$ intervals have positive and negative weight contributions, and the sign corresponds to the side from which we close the incident arcs.  With three $\op$'s, in order to disentangle the two types of contributions, we need something that breaks the left--right symmetry of the three operators and allows us to differentiate between the two methods of closing the arcs.  We can condition the left most arcs not to close on each other, which means fusion via $\Lmod_{\opa}$.  These three-point products have fusion $\op \{ \op \keto{\op} \}_{\opa}$. This set of operators are dominated by the logarithmic module $\Imod_{\poplog}$, with leading terms
\begin{equation}
\label{equ:Pi_3pt_Fusion}
\fl
\op(\ve) \left\{ \op(\l \ve) \keto{\op}\right\}_\opa = \Pi^\l_h \keto{\op}+\ve\frac{\sqrt{3}}{\pi} \F^\l_3 \left( \keto{\poplog} + \log (q_3\ve)  \keto{\pop}\right)
+ \ve\F^\l_2 \keto{\pop}
+\mathrm{O}(\ve^2)\; ,
\end{equation}
where $\Pi_h^\l$ is the Cardy/Smirnov crossing formula (we called it $Z_{\widehat{12}\cdot \widehat{34}}$ in an earlier section) and $q_3$ is an arbitrary constant.  The explicit forms are
\begin{eqnarray} \label{equ:Cardys_Form}
\Pi_h^\l&=\frac{3\Gamma(2/3)}{\Gamma(1/3){}^2}\l^{1/3}\,{}_2F_1(1/3,2/3;4/3;\l)\; ,\\
\F^\l_2 &=\frac{\Gamma(2/3)}{ 2\, \Gamma(1/3)^2}\l^{4/3}(1-\l)^{1/3}{}_3F_2\left( 2/3,1,1;2,7/3;\l\right)\; .
\end{eqnarray}

We have an explicit understanding of the configurations that contribute to the $\{\op\op\op\}_{\pop}$ set of operators.  Now we ask which what configurations correspond to the $\poplog$ operator in \eref{equ:Pi_3pt_Fusion})? When we combine this three operator set with another boundary operator at some distant point $x\gg\ve$, we get the same four point conformal block that Cardy used to derive the horizontal crossing probability
\be
\langle \op(x) \op (\ve) \left\{\op(\ve \l) \keto{\op} \right\}_\opa =
\Pi_h \left[\frac{\l (x-\ve)}{x-\l \ve}\right]\; .
\ee
The cross ratio varies with the location of $x$. Letting $x\to \infty$ we get the usual $\Pi_h^\l$.

The expansion of this crossing probability in powers of $\ve/x$,
\begin{equation*}
\Pi_h \left[\frac{\l(x-\ve)}{x-\l \ve}\right]=
\Pi_h^\l- \frac{\ve}{x} \l(1-\l)\partial_\l\Pi_h^\l+\mathrm{O}\left(\frac{\ve}{x}\right)^2\; ,
\end{equation*}
can be compared to the four-point function with the operator product expansion (\ref{equ:Pi_3pt_Fusion}), if we note that
\begin{equation}\langle \op(x) \keto{\op}=1\qquad
\langle \op(x) \keto{\poplog}=-(2x)^{-1}\; .
\end{equation}
Then it's easy to verify that the two expansions agree because
\begin{equation*}
\frac{\sqrt{3}}{\pi}\F^\l_3=2\l(1-\l)\partial_\l \Pi_h^\l\; .
\end{equation*}

In the limit $x\gg\ve$ the local operators are isolated enough that their internal properties uncouple from their interaction with distant operators, leading to the leading term: $\Pi_h^\l \keto{\op}$.  Dropping this term gives contributions where the $\op\{\op\op\}_\opasub$ differs from a $\op$ as a result of the extra arc anchored to the right of propagating arc.  Due to locality, the contractible arc is only significant if it touches some distant object and the contributions are dominated by three propagating paths, as was the case for the $\{\op\op\op\}_{\pop}$ operator set.

The configurations associated with the $\poplog$ differ from $\pop$ configurations in two important regards.  First the addition of the contractible arc is an \emph{active} modification of the correlation function and modifies the set of arc classes.  This means $\poplog$ cannot be added into an existing correlation function like a $\pop$ interval can.  Second, it breaks the left/right symmetry; as we've written it we consider a small extra arc explicitly on the right of the propagating arc.  Both of these considerations distinguish this logarithmic construction from it's regular counterpart.  In particular $\poplog$ only leads to positive weight contributions in correlation functions.

Now we'll to construct a $\poplog$ interval similar to $\Delta \op$.  It's our claim that we can follow the same procedure from the end of \sref{sec: 3ptFuse_Reg} to remove the internal structure of the three operators.  We ignore the contributions from the $\pop$ operator and divide by $2 \F_3^\l$ then integrate the leading weight one operator across an interval;
\be \label{equ: Log Op Int Int}
\frac{\sqrt{3}}{2\pi}\int_{x_1}^{x_2} \poplog(w)dw=:\Delta\Psi(x_1,x_2)\; .
\ee
Unlike the case with the $\pop$ integration there is no operator $\Psi$ such that $\partial\Psi=\poplog$, but we adopt this style of notation for convenience.
\begin{figure}[ht] 
\centering
\makebox[0.2 \columnwidth]{\raisebox{3ex}{$\Delta\Psi(x_1,x_2)\; :$}}
\includegraphics*[width=0.15\columnwidth]{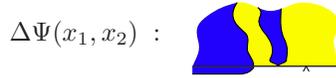}
 \caption{ \label{fig:poplog_Convention}
The $\Delta\Psi$ integral includes a distinct cluster anchored between the bounds of integration that propagates with a domain wall without touching it.  Multiple such clusters may exist for each cluster configuration.  In that case the configuration will contribute once for each suitable cluster.
}
\end{figure}
The integration of $\poplog$ gives weight for any arc along the edge, effectively creating a non--local arc attachment like we saw with $\Delta \op$.  However, the arc attached to the $\Delta\Psi$  is not merely a pre--existing arc originating somewhere else that brushes the boundary.  It is an additional arc, new to the configuration that must be considered when defining allowed arc classes.

Another feature important to $\Delta \Psi$, is that a single configuration may contribute to the integral multiple times if it has multiple arcs satisfying the properties of the configuration.  In particular we may actually find the expectation number of arcs instead of the probability that such arcs exists \cite{SimmonsKlebanZiffJPA07}.

\begin{figure}[ht] 
\centering
\makebox[0.2 \columnwidth]{\raisebox{5ex}{$\bra \Delta \Psi \Delta \op \ket\; :$}}
\includegraphics*[width=0.35\columnwidth]{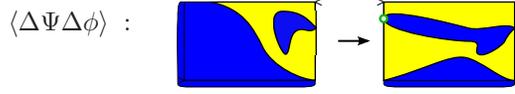}
 \caption{ \label{fig:Pi_h_nob_Convention}
The first step is to mark the extra arc attached to the $\Delta\Psi$ interval.  The second is to draw this loop in to interact with the passive $\Delta\op$ interval.  Since the loop attached to $\Delta \Psi$ is the only arc not attached to the $\op$ we're moving, this is the only contribution to this block.
}
\end{figure}

As an example consider this correlation function, which utilizes the $\Delta \Psi$ interval:
\begin{eqnarray}\nonumber
\bra \Delta \Psi(x_3,x_4) \Delta \op(x_1,x_2) \ket&=\frac{\sqrt{3}}{2\pi}\int_{x_3}^{x_4}\!\! \bra\poplog(w)\Delta\op(x_1,x_2) \ket dw\\ \label{equ: DYDf_CorFunc}
&=\frac{\sqrt{3}}{4\pi}\log\left[ \frac{(x_4-x_2)(x_3-x_1)}{(x_4-x_1)(x_3-x_2)}\right]\; .
\end{eqnarray}
To construct the arc configurations that contribute to this correlation function we first write down all of the arcs implied by the fusion channels of the conformal blocks, including single $\op$ operators on the left hand side of any $\Delta\Psi$ or $\Delta\op$ intervals for now.  We then include a `bubble' along the integration path of the $\Delta \Psi$ interval to represent the contractible internal arc associated to the $\poplog$ operator (left side of \fref{fig:Pi_h_nob_Convention}).
Finally we shift the $\op$, creating the $\Delta \op$, and keep only contributions where other arcs intersect between the two shift points, paying careful attention to the $\Delta \op$ sign convention  (right side of \fref{fig:Pi_h_nob_Convention}). This result is conformally invariant.

Because of the integration of $\Delta \Psi$ we will get the expected number of clusters satisfying this arc configuration.  The blue cluster in \fref{fig:Pi_h_nob_Convention} is a crossing cluster that doesn't touch the bottom edge, so we've calculated the expected number of horizontal crossing clusters that don't touch the bottom of the rectangle \cite{SimmonsKlebanZiffJPA07}.

We'll run into issues with $\Delta \Psi$ since $\poplog$ couples to $\op$ and $\pop$.  Under a general conformal mapping $z \mapsto u=f(z)$ the logarithmic operator transforms as
\begin{equation*}
\poplog(z) \mapsto -\frac{u''}{4 u'} \op\left(u\right)+u' \log u' \pop\left(u\right)+ u'\psi\left(u\right)\; .
\end{equation*}
After a mapping the ratio of lengths between any three operators is bound to change, including the internal ratio $\l$ around which we built our three operator fusion.
The first term in the transformation rule corrects for this local change,
\begin{equation*}
\Pi_h \left[\frac{f(\l \ve)-f(0)}{f(\ve)-f(0)}\right]=\Pi_h^\l-\frac{f''(0)}{4 f'(0)}\ve \frac{\sqrt{3}}{\pi}\F^\l_3+{\rm O}(\ve^2),
\end{equation*}
but this coupling means that we cannot construct an expression for $\Delta \Psi$ that is fully invariant under conformal transformations.  Since the troublesome terms always couple to the $\op$ operator, we can still get a  unique conformally invariant correlation function, whenever $\Delta \Psi$ couples to the $\Lmod_{\pop}$ module as in the example above.

\Eref{equ: DYDf_CorFunc} is conformally invariant, but if we try to couple $\Delta \Psi$ to the plain $\op$ we get an expression that is not.
Under a conformal mapping $u=f(x)$ this correlation function transforms as
\begin{equation*}
\fl
\bra \Delta \Psi\left(x_3,x_4\right)\op\left(x_1\right) \ket=\frac{\sqrt3}{4\pi} \log\left[\frac{x_4-x_1}{x_3-x_1}\right]
=-\frac{\sqrt{3}}{8\pi}\log\left[\frac{f'(x_4)}{f'(x_3)}\right]
+\bra \Delta \Psi\left(u_3,u_4\right)\op\left(u_1\right) \ket\; .
\end{equation*}
We may attempt to associate this correlation function with configurations like the left hand side of figure \ref{fig:Pi_h_nob_Convention}.  However, the issue with this correlation function stem from the regularization issues that face fully contractible loops.  In particular every configuration has an infinite number of loops attached to n arbitrary edge. To avoid these sorts of issues we'll always be sure to couple the $\Delta \Psi$ interval with a $\pop$ channel interval.

We identify $\poplog$ objects within the CG formalism using the three point operator set.  It's easiest to argue if we start with the four point vertex operator expression that gives the crossing probability between the intervals $(x_1,x_2)$ and $(x_3,x_4)$ and take the derivative with respect to $x_4$,
\begin{equation*}
\fl
\partial_{x_4}\bra \vn{1} \vn{2} \vn{3} Q_{-}^{(x_3,x_4)} \ket
= \bra \vn{1} \vn{2} \vn{3} \scn{4}\mathcal{V}_-(x_4)
\ket
\end{equation*}
The three operators at $x_1$, $x_2$ and $x_3$ fuse to $\poplog$ according to (\ref{equ:Pi_3pt_Fusion}), so we associate the logarithmic operator to $3\times \a_{1,2}^-=3/\sqrt{6}=\a_{1,4}^-=\a_+$,
\begin{equation*}
\frac12 \mathcal{V}_\psi^-:=\frac12\scp{}= \sim \poplog(x)\quad {\rm and}\quad \Delta\Psi(x_1,x_2)=\frac{\sqrt{3}}{4\pi}Q_+^{(x_1,x_2)}\; .
\end{equation*}
We include the one half so that the two point function $\langle \pop(x) \keto{\poplog}=1/(2x^2)$ is properly normalized.  While this operator is easily represented as a positive screening charge, this interpretation doesn't mesh with our understanding of the $O(n)$ loop conformal blocks and we should think of this as an integral of a three leg operator.  It's easy to verify that this convention works to give us the correlation function
\begin{equation*}
\bra \Delta \Psi(x_3,x_4) \Delta \op(x_1,x_2) \ket=\frac{\sqrt{3}}{4\pi}\int_{x_3}^{x_4}\hspace{-2ex}dw\!\!\int_{x_1}^{x_2} \hspace{-2ex}dz \bra \mathcal{V}_{\psi}^-(w) \mathcal{V}_{\pop}^+(z) \ket\; .
\end{equation*}
with precisely the right normalization.

\subsection{The stress tensor partner interval \label{sec: STPI}}

The interval we identify with the stress tensor partner is based on the $\Delta \Psi$ constructed in the last section.  Most of the details of this construction closely mirror the details of the $\Delta \Psi$ construction, so we'll be able to make this section quite brief.

First note that combining the $\Delta\op$ interval with another boundary change operator so that the total fusion channel is $\M_1$ gives us
\begin{eqnarray}
\{\Delta\op(x_1,x_2)\op(x_1)\}_1&=\{\op(x_2)\op(x_1)\}_1-1\\
&=\{\op(x_2)\op(x_1)\}_T\; .
\end{eqnarray}
This is an alternative formulation of the stress tensor interval.

\begin{figure}[ht] 
\centering
\makebox[0.15 \columnwidth]{\raisebox{2ex}{$\left\{\Psi \op\right\}_\tau\; :$}}
\includegraphics*[width=0.15\columnwidth]{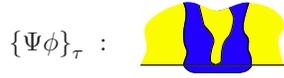}
 \caption{ \label{fig:tau_Convention}
This configuration has a non-local collection of loops attached to the interval, otherwise it behaves like an identity channel interval.}
\end{figure}

The analogous interval with the logarithmic counterpart is
\begin{equation*}
\{\Delta\Psi(x_1,x_2)\op(x_1)\}_\tau=\frac{\sqrt{3}}{2\pi}\int_{x_1}^{x_2} \{\poplog(w)\op(x_1)\}_\tau dw\; ,
\end{equation*}
which we'll denote with $\{\Psi(x_2)\op(x_1)\}_\tau$ and call the stress tensor partner interval.
This interval is interpreted in the natural way based on it's construction from a  $\Delta\Psi$  and a $\op$, and we represent is as a pair of boundary change operators with another loop attached somewhere in between (see \fref{fig:tau_Convention}).

As with the $\Delta \Psi$ interval we need to be careful how we associate this interval to arc configurations.  In particular this interval should always couple to an object in the stress tensor module, because the $\tau$ operator does not transform as a primary operator, and otherwise we'd get additional contributions from the identity channel under a conformal mapping.  Also we need to realize that due to the integration we'll get contributions form any loop that satisfies the specified geometry giving us expectation numbers of such arcs.

\begin{figure}[ht] 
\centering
\makebox[0.275 \columnwidth]{\raisebox{5ex}{$\bra \{ \Psi \op\}_\tau \{\op \op\}_T \ket \; :$}}
\includegraphics*[width=0.35\columnwidth]{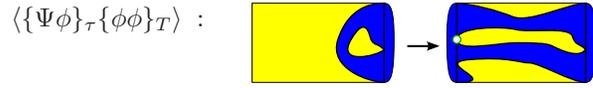}
 \caption{ \label{fig:Expected No-1_Convention}
The first configuration represents the $\Delta \Psi$ interval, the second represents the interaction of this interval with the stress tensor interval.}
\end{figure}

The simplest construction involving the stress tensor partner interval is the correlation function
\begin{equation*}
\bra \left\{\Psi(x_4)\op(x_3) \right\}_\tau \left\{\op(x_2)\op(x_1)\right\}_T \ket\; .
\end{equation*}
The arc schematic for these configurations, shown in \fref{fig:Expected No-1_Convention},
are found by drawing all allowed arc configurations, while temporarily leaving stress tensor intervals unmarked and treating the $\{\Psi\op\}_\tau$ interval as in identity with a small arc stuck to it, as in the left side of \fref{fig:Expected No-1_Convention}.  The we account for the stress tensor interval, by pulling in pars of arcs according to the sign convention in \fref{fig:Perc_TInt}.

This gives the expected number of \emph{extra} crossing clusters because the yellow loop in the figure could represent any space in between two crossing clusters \cite{SimmonsKlebanZiffJPA07}.  If there are $N$ crossing clusters in a configuration then the configuration contributes to the integral $N-1$ times, prompting our interpretation.

The Coulomb gas representation of this interval follows directly from the $\Delta \Psi$ representation,
\begin{equation*}
\{ \Psi(x_4) \op(x_3)\}_\tau \sim \frac{\sqrt{3}}{4\pi}\int_{x_3}^{x_4}\!\!\!\! du\; \mathcal{V}_{\poplog}^-(u)\vn3\; ,
\end{equation*}
but we can't couple it directly to the charge neutral representation of \STI~since there is a charge mismatch.  Instead we construct another representation of the stress tensor interval
\begin{equation*}
\{ \op(x_2) \op(x_1)\}_T \sim K\int_{x_1}^{x_2}\!\!\!\! dw\; \mathcal{V}_{\pop}^+(w) Q_-^{(w,x_1)} \vn{1}
\end{equation*}
that is built of an integrated $\pop$ operator to produce the right change.

Now that we've taken the time to construct these intervals within the Coulomb gas formalism, we can modify them by adding charge neutral identity intervals.  In the next section we'll show how the process can be used to make calculations using the example of the six point function.

\section{ \label{sec: 3 Interval Xing}
Percolation crossing probabilities in conformal hexagons}

\begin{figure}[htb]
\centering
\includegraphics[width=0.7\columnwidth]{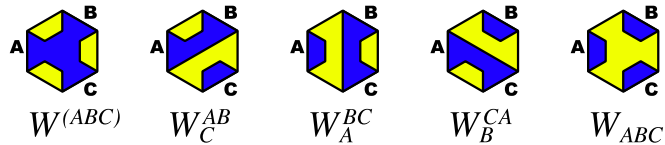}           
\caption{
\label{fig: Hex Arc Configurations} 
The five possible arc configurations associated with $6$SLE$_6$.
}
\end{figure}

In ref.~\cite{Dubedat2006} Dub\'edat analyzed the $6$SLE$_\k$ process, which for $\k=6$ can be used to describe percolation in a hexagon with yellow and blue boundary conditions on alternating sides.  This process corresponds to the correlation function $\bra \op \op \op \op \op \op \ket$ with the operators situated at the corners of the hexagon. The boundary conditions allow us to decompose the percolation configurations into five classes of arc configurations  depending on the connectivity of the three wired intervals (see \fref{fig: Hex Arc Configurations}).

Dub\'edat worked with the systems of six differential equations that arise form the SLE processes at each point and with the additional three differential equations implied by conformal symmetry.  This lead him to an integral expression encompassing the different conformal blocks of the function depending on the choice of integration paths.  Dub\'edat's integral expressions are equivalent to the Coulomb gas expressions we discuss, though the integration paths differ, leading to different conformal blocks.

As usual, we'll take advantage of the conformal invariance of these crossing probabilities and simplify the problem for ourselves by considering the problems in the upper half plane.  It remains easier to envision the problem in the hexagon, so we'll continue to use that geometry in our illustrations.   For concreteness let's assume that operators $\op_i:=\op(x_i)$ mark the boundaries of three blue intervals $(x_1,x_2)$, $(x_3,x_4)$ and $(x_5,x_6)$, labelled $C,B$ and $A$ respectively.

Based on the discussion of $O(n)$ fusion rules we can easily construct Coulomb gas expressions for conformal blocks and identify which of the five weights contribute to each.  If we have each blue interval fuse to the identity then we get the correlation function
\begin{eqnarray} \label{Triple_Identity_Correlation}
\fl
\left\langle \left\{\phi_6 \phi_5\right\}_{1} \left\{ \phi_4 \phi_3 \right\}_{1} \left\{\phi_2 \phi_1 \right\}_{1} \right\rangle=W^{(ABC)}+W^{AB}_C+W^{BC}_A+W^{CA}_B+W_{ABC}
\nn
\fl
\hspace{2cm}
=K^2\bra\vp{6} \vn{5} \vn{4} Q_-^{(x_3,x_4)} \vn{3}\vn{2}Q_-^{(x_1,x_2)}\vn{1}\ket\; ,
\end{eqnarray}
recall the $K=\Gamma(2/3)/\Gamma(1/3)^2$ for $\k=6$.  Using Mathematica to numerically evaluate this integral for several choices of $x_i$ gives $1$ in each case, as we'd expect from the percolation interpretation and fusion rules.

Alternately we consider the contributions from the conformal block
\begin{eqnarray} \label{Double_opa_Correlation}
\fl
\left\langle \left\{\phi_6 \phi_5\right\}_{\opa} \left\{ \phi_4 \phi_3 \right\}_{\opa} \left\{\phi_2 \phi_1 \right\}_{1} \right\rangle=W^{(ABC)}+W^{AB}_C\nn
\fl
\hspace{2cm}
=K^2\bra\vp{6}Q_-^{(x_5,x_6)} \vn{5} \vn{4} \vn{3}\vn{2}Q_-^{(x_1,x_2)}\vn{1}\ket\; .
\end{eqnarray}
This collection of operators could be built sequentially, starting with the pair of two leg intervals (see the first line of \fref{fig:4pt_opa_blocks}) that require intervals $A$ and $B$ to share a bulk crossing cluster.   Adding the charge neutral identity interval then colors interval $C$ blue, which doesn't effect existing crossing clusters, but does allows connections from $A$ to $B$ made through blue boundary sites on $C$.  It's easy to find the combinations $W^{(ABC)}+W^{CA}_B$ and $W^{(ABC)}+W^{BC}_A$ using a similar conformal block where the identity interval sits on $B$ or $A$ respectively.

We can write three similar expressions by swapping to the yellow intervals instead of the blue.  This yields $W_{ABC}+W^{AB}_C$, $W_{ABC}+W^{CA}_B$, and $W_{ABC}+W^{BC}_A$, giving us a total of seven conformal blocks.  From these we isolate any of the five arc weights we like, using only the simplest $O(n)$ model boundary conditions and real integrals.

Dub\'edat reported closed expressions for the arc configurations in \fref{fig: Hex Arc Configurations} in the highly symmetric case of the regular hexagon when $W^{BC}_A=W^{CA}_B=W^{AB}_C$ and $W^{(ABC)}=W_{ABC}$. In this case the particular result is
\begin{eqnarray} \label{equ:D_disjoint}
\overline W^{BC}_A=\overline W^{CA}_B=\overline W^{AB}_C = \frac{3^{3/2}\, \Gamma(2/3)^9}{2^{7/3}\, \pi^5} {}_3F_2\left(1,\frac{5}{6},\frac{5}{6};\frac{3}{2},\frac{3}{2}\bigg|1\right)\\ \label{equ:D_joined}
\overline W^{(ABC)}=\overline W_{ABC}=\frac12-\frac{3^{5/2}\, \Gamma(2/3)^9}{2^{10/3}\, \pi^5} {}_3F_2\left(1,\frac{5}{6},\frac{5}{6};\frac{3}{2},\frac{3}{2}\bigg|1\right)\; .
\end{eqnarray}
Using Mathematica to numerically evaluate the integral in \eref{Double_opa_Correlation} produces these values.


\subsection{\label{sec: Decomposed Xings}
Cluster configurations with a free boundary}

Now we ask how these results can be used to analyze the crossing probabilities of blue clusters in the hexagon without any blue boundary segments.  From the point of view of the blue clusters this corresponds to free boundary conditions as our choice of boundary conditions doesn't change the connectivity of the bulk blue clusters.

\begin{figure}[ht]
\centering
\includegraphics[width=0.7\columnwidth]{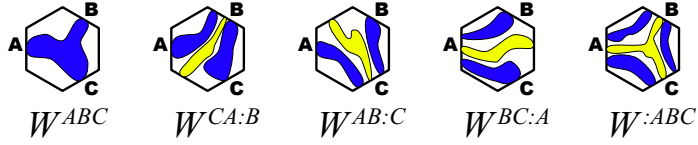}
\caption{
\label{fig: Hex Xing Decomp} 
We can further decompose the configurations that contribute to $W^{(ABC)}$  into these five subclasses in the presence of a free boundary.
}
\end{figure}

Of the five configurations shown in \fref{fig: Hex Arc Configurations}, four retain their meaning even if we swap to a homogeneous yellow boundary.  The case we need to look at more closely is when all three blue intervals are connected: $W^{(ABC)}$.  When the boundary is free we see that the configurations that contributed to this weight further decompose into five sub-configurations as shown in \fref{fig: Hex Xing Decomp} .

We may ask to find the probability that all three of the sides $A$, $B$ and $C$ are connected by a common cluster, corresponding to $W^{ABC}$.  The issue is that there are several ways for clusters to connect these sides involving disjoint clusters, and the alternating blue and yellow boundary conditions don't distinguish between them.

Consider the case where a cluster from $A$ connects to $C$ and a distinct cluster from $B$ connects to $C$ as well.  This configuration is represented by $W^{AB:C}$ and it contributes to the weight $W^{(ABC)}$ when the blue boundary conditions are present.  We can likewise define $W^{CA:B}$ and $W^{BC:A}$  when the disjoint clusters both hit $B$ and $A$ respectively.

\begin{figure}[ht]
\centering
\includegraphics[width=0.7\columnwidth] {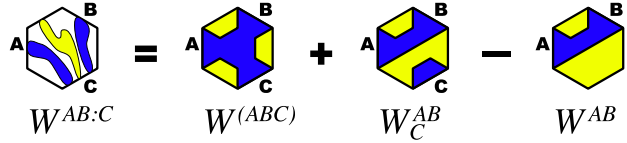}           
\caption{
\label{fig: Hex Arith1} 
We consider the wired interval $C$, and the effect it's inclusion has on crossing events between intervals $A$ and $B$.  This allows us to identify the weight $W^{AB:C}$ in terms of $6$ and $4$ arc weights.
}
\end{figure}

This case, where two clusters are considered joined because they interact with a fixed boundary segment, corresponds precisely to the stress tensor interval.  The configurations decompose as in \fref{fig: Hex Arith1}
where $W^{AB}$ is the probability of a crossing cluster between $A$ and $B$, given by $\Pi_h^\l$ for the right choice of cross-ratio.  We alluded to this above when we talked of adding the identity channel interval to \eref{Double_opa_Correlation}.  If we make that explicit by subtracting the identity operator from the neutral identity channel we get the expression
\begin{eqnarray} \label{Double_opa_T_Correlation}
\fl
\left\langle \left\{\phi_6 \phi_5\right\}_{\opa} \left\{ \phi_4 \phi_3 \right\}_{\opa} \left\{\phi_2 \phi_1 \right\}_{T} \right\rangle=W^{(ABC)}+W^{AB}_C-W^{AB}=W^{AB:C}\nn
\fl
\hspace{2cm}
=K^2\bra\vp{6}Q_-^{(x_5,x_6)} \vn{5} \vn{4} \vn{3}\vn{2}Q_-^{(x_1,x_2)}\vn{1}\ket\nn
\fl
\hspace{3cm}
- K\bra\vp{6}Q_-^{(x_5,x_6)} \vn{5} \vn{4} \vn{3}\ket\; .
\end{eqnarray}
Of course we can cycle through $A$, $B$ and $C$ to find $W^{CA:B}$ and $W^{BC:A}$ as well.

For the regular hexagon we can write an expression for this weight using \eref{equ:D_disjoint} and \eref{equ:Cardys_Form} with the appropriate cross ratio $\l=1/3$
\begin{equation*}\fl
\overline W^{AB:C}=
\frac12-\frac{3^{3/2}\, \Gamma(2/3)^9}{2^{10/3}\, \pi^5} {}_3F_2\left(1,\frac{5}{6},\frac{5}{6};\frac{3}{2},\frac{3}{2}\bigg|1\right)-\frac{3^{2/3}\Gamma(2/3)}{\Gamma(1/3){}^2}\,{}_2F_1\left(\frac13,\frac23;\frac43\bigg|\frac13\right)
\end{equation*}

So far in this section we've gotten away without using any logarithmic operators in these correlation functions.  This has been possible because the cluster configurations we've considered can be defined using only the three arcs generated by the six boundary change operators.  In the case of the $W^{AB:C}$ configurations we just finished with, we needed to know whether one of these three arcs was incident on the boundary, but this remained a passive observation that didn't modify the arc classes we considered.

The final configurations that we need to separate are $W^{ABC}$ and $W^{:ABC}$.  These configurations \emph{cannot} be distinguishes using only the outer arcs of the blue intervals.  The difference between these configurations hinges on whether a yellow cluster spans between $A$, $B$ and $C$ without hitting any of the yellow boundary intervals, but the arc that marks the edge of this internal cluster isn't connected to the boundary change operators in the corners.  Non-local arcs like this originate from logarithmic operators.

To identify this contribution we recall that the logarithmic correlation function $\langle\{\Psi_6 \op_5\}_\tau \{\op_4 \op_3\}_T \rangle$ represents the expected number of extra crossing clusters that span between intervals $A$ and $B$. Now coloring the interval $(x_1,x_2)$ blue as well doesn't change the number of crossing clusters from $A$ to $B$; it doesn't matter whether the cluster adjacent to $C$ touches the interval or not it still crosses.  However we will gain a weight from configurations associated with the weight $W^{:ABC}$ because coloring $C$ blue connects any two clusters incident on $C$, and for the $W^{:ABC}$ configurations this creates another crossing cluster from $A$ to $B$, increasing our expectation value. Since coloring $C$ increase the number of crossing clusters by one cluster for each configuration in $W^{:ABC}$ the correlation function for the expected number of extra crossing clusters with a stress tensor channel interval added at $C$ gives precisely this weight:
\begin{eqnarray} \fl \label{equ:Tau_T-T_Correlation}
\left\langle \left\{\Psi_6 \op_5 \right\}_\tau \left\{ \op_4 \op_3\right\}_T\left\{\phi_2 \phi_1 \right\}_T\right\rangle = W^{:ABC}\nn \fl
\;\;\;= \frac{\sqrt{3} K^2}{4\pi}\int_{x_5}^{x_6}\!\!\!\! du  \int_{x_3}^{x_4}\!\!\!\! dw\;\; \bra\mathcal{V}_{\poplog}^-(u)\vn5 \mathcal{V}_{\pop}^+(w) Q_-^{(x_3,w)} \vn{3} \vn{2}Q_-^{(x_1,x_2)} \vn{1}\ket\nn
- \frac{\sqrt{3} K}{4\pi}\int_{x_5}^{x_6}\!\!\!\! du  \int_{x_3}^{x_4}\!\!\!\! dw\;\; \bra\mathcal{V}_{\poplog}^-(u)\vn5 \mathcal{V}_{\pop}^+(w) Q_-^{(x_3,w)} \vn{3}\ket
\end{eqnarray}
While this is a relatively unwieldy quadruple integral involving twenty two non-trivial factors, it gives the probability that three separate clusters connect $A$, $B$ and $C$ as in \fref{fig: Hex Xing Decomp}.  Now that we have an expression for this crossing probability we can isolate the free boundary crossing probability if we're so inclined.  We'll finish by writing down the numerical values for the various weights in the case of the regular hexagon
\begin{eqnarray*}
\overline W^{BC}_A=\overline W^{CA}_B=\overline W^{AB}_C=0.1408552\ldots\\ 
\overline W_{ABC}=0.2887170\ldots\\
\overline W^{AB:C}=\overline W^{AB:C}=\overline W^{AB:C}=0.0108875\ldots\\
\overline W^{\,:ABC}=0.000364657\ldots\\
\overline W^{ABC}=0.2556897\ldots\;.
\end{eqnarray*}
There are closed form expressions for most of these weights, with only $W^{ABC}$ and $W^{:ABC}$ depending on the numerical integration of \eref{equ:Tau_T-T_Correlation}, for which we used Mathematica.

\section*{Conclusion}

In this paper we consider the $c=0$ logarithmic CFT generated by the $\phi_{1,2}$ operator.  We focused on a correspondence between two-dimensional critical percolation and seven operators in the theory.

In section \ref{sec: O n Loop} we discuss the $O(n)$ loop model in the presence of boundary operators $\op$ that create open arcs attached to the boundary.  We identify classes of configurations based on the connectivities of these arcs, and devise a scheme to determine the weight given to each arc class in a particular conformal block by associating boundary conditions in correspondence with the fusion channels present.  The boundary conditions distinguish identity channel intervals, pairs of $\op$ that fuse via the identity module allowing either contractible or propagating arcs, and two-leg channel intervals, pairs that fuse via the $\opa$ module that only allow propagating arcs.

We review the Coulomb gas method, discussing how to identify fusion channels from vertex operators and screening charges, which allows us to write conformal blocks with specific fusion properties, and in turn identify the weighted contribution from each arc class.  We emphasized the identity channel interval, discussing explicitly how these intervals can be added to existing Coulomb gas conformal blocks via a  properly normalized charge neutral construction without modifying the existing fusion channels.

In section \ref{sec:Perc} we discussed the logarithmic $c=0$ conformal field theory specific to critical percolation.  Locality  played an important role in this discussion since changes to the boundary conditions don't effectively change the weights of bulk configurations.  After briefly reviewing the relevant operators and modules we proceeded to identify the percolation configurations associated to a variety of intervals designed to yield these operators in the fusion that results from shrinking them to a point.   The two staggered logarithmic modules $\Imod_{1,4}$ and $\Imod_{1,5}$ are of particular interest.

The interval $\Delta \op$ is the net change from moving the $\op$ from the left to right edge of a boundary interval.  This change isolates configurations were other arcs interact with the boundary in question either before or after the move, giving a positive weight when an arc hits from the left after the move, and a negative sign when they hit on the right before the move.  This operator is passive, only feeling the effect of the existing arcs.

We contrast $\pop$ with logarithmic partner $\poplog$. We argue that $\poplog$ is non-passive, introducing a contractible loop on one side of the $\op$ to which it's attached.  Using the three point fusion $\op^3=\Imod_\psi$ that we derive in the appendix, we argue that the operator can be integrated to creating an interval that effectively has a non-local contractible loop attached to it.  This interval fuses to $\poplog$ and we call it $\Delta \Psi$.  However,  $\Delta \Psi$  is not inherently conformally invariant due to the transformation rules for $\poplog$ that introduce the operator $\op$, and we must restrict ourselves to coupling $\Delta \Psi$ with $\Delta \op$ to retain conformal invariance.  This forces the contractible loop on the interval to interact with the boundary elsewhere, which removes the transformation and scaling issues related to the regularization of a loop with no fixed size.

The stress tensor interval $\{\op\op\}_T$ is associated to configurations whose contribution to the conformal block change depending on the boundary conditions on the interval.  This requires at least two disjoint clusters incident on the interval.  If fusion channels require the disjoint clusters to connect then the configuration contributes a positive weight.  If the other fusion channels require the clusters to remain disjoint then the configuration gets a negative weight. This is another passive effect, since it's dependent on existing arcs and has both positive an negative contributions.  We can isolate the stress tensor interval in Coulomb gas formalism by taking the difference of two correlation functions one with an identity channel interval, one without.

We construct an interval $\{\Psi \op\}_{\Tlog}$ for the logarithmic partner to the stress tensor as well, building upon the idea of the $\Delta \Psi$ by combining this integral with an other $\op$.  The resulting interval has a small contractible look attached to a fixed boundary section that has arcs attached at either end.  This operator is not conformally invariant unless it is attached to a stress tensor interval.

We end by using these intervals in various combinations to calculate crossing probabilities in hexagons with free boundaries.  We end by establishing new integral expressions for the probability that a percolation cluster touches every third side of the hexagon.  This requires combining several conformal blocks including one with one $\{\Psi\op\}_{\Tlog}$ and two $\{\op\op\}_T$.

The requirement that the logarithmic intervals we've constructed must couple to their regular counterparts is a weakness that we'd like to see resolved in future work.

\section*{Acknowledgments}
We would like to thank S.~Flores for productive conversations. We would also like to thank the organizers of \emph{Advances in Percolation and Related Topics}, U.~Michigan 2012, and \emph{Conformal Invariance, Discrete Holomorphicity and Integrability}, U.~Helsinki 2012 where parts of this work were presented and developed.   We would finally like to thank the Maine Maritime Academy professional development fund for travel support.  

\appendix{}

\section{Three point fusion \label{Appendix}}

We begin with the three point fusion and the Ansatz
\begin{eqnarray}\label{ansatz}  \fl
\cs &:= \op(\ve) \op(\ve \l) \keto{\op}
=A^{\l,\ve} \keto{\op}
+ \sum_{\{ k\}^* }B_{\{k\}}^{\l,\ve} L_{-\{k\}}\keto{\poplog}+\sum_{\{ k\}^{*} }C_{\{\ell\}}^{\l,\ve} L_{-\{\ell\}}\keto{\pop}\; ,
\end{eqnarray}
where $\{ k \}^*$ are the non-increasing partitions of the natural numbers.  The asterisk indicates that we remove a partition for each null descendant of $\pop$, so that the $L_{-\{k\}^*}\keto{\pop}$ remain linearly independent.  The action of the Virasoro Generators upon $\cs$ can be deduced using the commutation relation $[L_n,\op(x)]=x^{n+1}\pop$.
\begin{equation*}
L_{n}\op(x)\op(y)\keto{\op}=\op(x)\op(y)L_{n}\keto{\op}+\left(x^{n+1}\p_x+y^{n+1}\p_y\right)\op(x)\op(y)\keto{\op}
\end{equation*}
If we make the change of variables $\{x,y\}= \{ \ve, \ve \l\}$ then this relation becomes
\be \label{equ:V_Gen_Action}
L_{n}\cs=\op(\ve)\op(\ve \l)L_{n}\keto{\op}
+\Do{n} \cs\; ,
\ee
where we have defined the differential operator
\begin{equation*}
\Do{n}:=\ve^n\left(\ve \p_\ve-\l(1-\l^n) \p_\l\right)\; .
\end{equation*}
If we compare this action with that resulting from the direct application of Stress tensor modes to the right hand side of equation \ref{ansatz} we can deduce the forms of low level coefficients.
We recall the action of the Virasoro generators
\begin{equation*}
\begin{array}{l@{\,}ll@{\,}ll@{\,}l}
L_1 \keto{\op}&=0 &L_1 \keto{\pop}&=0&L_1 \keto{\poplog}&=\a\keto{\op}\\
 L_0 \keto{\op}&=0&L_0 \keto{\pop}&=\keto{\pop}&L_0 \keto{\poplog}&=\keto{\poplog}+\keto{\pop}
\end{array}
\end{equation*}
where the constant $\a=-1/2$ is known determine it independently.

The action of the scaling generator $L_0$ yields the following equations
\begin{eqnarray*}
\fl
\ve \p_\ve A^{\l,\ve}= 0 &\Rightarrow\quad A^{\l,\ve}= A^\l\\ \fl
\ve \p_\ve B_{\{k\}}^{\l,\ve}= B_{\{k\}}^{\l,\ve} (1+|k|)&\Rightarrow\quad B_{\{k\}}^{\l,\ve}= \ve^{|k|+1} B_{\{k\}}^\l\\ \fl
\ve \p_\ve C_{\{ k\}}^{\l,\ve}= B_{\{k\}}^{\l,\ve} +C_{\{ k \}}^{\l,\ve} (1+|k|)\quad&\Rightarrow\quad C_{\{ k\}}^{\l,\ve}= \ve^{|k|+1} \left(B_{\{k\}}^\l \log \ve  + C_{\{ k \}}^\l\right)\; .
\end{eqnarray*}

We determine the leading coefficients using the lower order Virasoro generators and the null state conditions for the three $\op$ operators. The terms of interest are,
\be \label{modansatz} \fl
\cs =A^\l \keto{\op}
+ \ve \left( B_0^\l \xs+C_0^\l  \keto{\pop} \right)
+ \ve^{2}L_{-1} \left( B_1^\l \xs+C_1^\l \keto{\pop} \right)
 + \mathrm{O}\left(\ve^3\right)\; ,
\ee
where we simplified our notation by defining
\begin{equation*}
\xs:=\keto{\poplog}+\log \ve \keto{\pop} \qquad \textrm{as well as}\qquad\Del{n}{m}:=m-\l(1-\l^n)\p_\l\; .
\end{equation*}
a differential operator in $\l$ so that
\begin{eqnarray*}
\Do{n} \ve^m F(\l) L_{\{k\}}\keto{\op} &=\ve^{n+m} \Del{n}{m}F(\l)  L_{\{k\}}\keto{\op}\\
\Do{n} \ve^m F(\l)  L_{\{k\}}\xs &=\ve^{n+m} \Del{n}{m}F(\l)  L_{\{k\}}\xs+\ve^{n+m} F(\l)  L_{\{k\}}\keto{\pop}\; .
\end{eqnarray*}

Letting $n=1$ in equation (\ref{equ:V_Gen_Action}) leads to the differential equations
\begin{eqnarray} \label{equ:eq1}
\a B_0^\l = \Del{1}{0}A^\l\\ \label{equ:eq2}
2 B_1^\l =\Del{1}{1}B_0^\l\nn
2C_1^\l+(2+\a)B_1^\l =\Del{1}{1} C_0^\l+B_0^\l
\end{eqnarray}
and letting $n=2$ gives the further differential equation
\begin{eqnarray} \label{equ:eq3}
3 \a B_1^\l&=\Del{2}{0} A^\l
\;.
\end{eqnarray}

These differential equations provide us with a great deal of information about the functional coefficients.  For example equations (\ref{equ:eq1}, \ref{equ:eq2} and \ref{equ:eq3}) can be used to show that
\begin{eqnarray*}
0&=\left(3\Del{1}{1}\Del{1}{0}-2\Del{2}{0}\right) A^\l\\
&=\l(1-\l)\left[3\l(1-\l)\p_\l{}^2+2(1-2\l)\p_\l\right]A^\l\; ,
\end{eqnarray*}
which is the differential equation for the horizontal crossing formula and allows two solutions: $1$ and Cardy's result for the horizontal crossing probability $\Pi_h^\l$, see equation (\ref{equ:Cardys_Form}).

If we write
\begin{equation*}
A^\l=k_1-3\, \a \, k_2 \l^{1/3} {}_2F_1\left(1/3,2/3;4/3;\l \right)
\end{equation*}
then the above differential equations imply the following forms for the remaining coefficients
\begin{eqnarray*}
B_0^\l = k_2 \left[ \l(1-\l) \right]^{1/3}&
B_1^\l =\frac{1+\l}{3}B_0^\l\\
C_1^\l =\frac{1-2\l-\a(1+\l)}{6}B_0^\l+\frac{1}{2}\Del{1}{1}C_0^\l
\end{eqnarray*}

We have already gleaned a great deal of information about the OPE coefficients without manipulating the null states.  It's interesting that the ansatz that we started with was sufficient to determine the  a four point function.   We can derive all of the coefficients except $C_0^\l$ and $\a$, but we may ask if these can be determined by using the null state conditions.

The null state condition for $\op(0)$ implies that
\begin{eqnarray}
0
&= \op(x) \op(y) \left[2L_{-2}-3L_{-1}{}^2\right] \keto{\op}\nn
&= \left[2\left(L_{-2}-x^{-1}\p_x-y^{-1}\p_y\right)-3\left(L_{-1}-\p_x-\p_y\right){}^2\right]\op(x) \op(y) \keto{\op}
\end{eqnarray}
where the second line is the result of propagating the generators to the exterior of the fusion.  
Translating this into the usual coordinates gives
\begin{equation*}
0 =\Big[ \left(2L_{-2}-3L_{-1}{}^2\right) +6\Do{-1} L_{-1}
-\left(
2\Do{-2} +3\Do{-1}{}^2
\right)\Big]\cs\; .
\end{equation*}
Applying this equation to (\ref{modansatz}) and rearranging terms gives us several redundant relations and one new relation
\be \label{equ:eq7}
0=\Eq_0:=\left(3\Del{-1}{0} \Del{-1}{1}+2\Del{-2}{1}\right)C_0^\l+k_2 \frac{(1-\l)^{1/3}(2+6\a+\l)}{\l^{2/3}}\; .
\ee

Alternately we can examine the null state of the operators that sit away from the origin. We translate the desired operators in question onto the origin, take advantage of the null-state, and then translate then back. 
We take the general null state equation
\begin{eqnarray}
\fl
0&=\E^{-y L_{-1}} \op(x)\op(y)\left[2 L_{-2}-3L_{-1}{}^2 \right] \keto{\op}\nn \fl
&=\Bigg[ 2 \sum_{s=-1}^\infty (-y)^s L_{-(s+2)} -2\left(x^{-1}\p_x+y^{-1}\p_y\right)
-3\left(\p_x+\p_y\right)^2\Bigg]\op(x-y)\op(-y)\keto{\op}
\end{eqnarray}
where the null state of the operator now at $-y$ has been exploited.

If we change to our preferred coordinates via $\{x,y\}=\{\ve(1-\l),- \l \ve \}$ then the null state condition that comes from $\op (\l \ve)$ implies that
\begin{equation*}
0=\Bigg[ 2 \left( \sum_{s=-1}^\infty \ve^s \l^s L_{-(s+2)}\right) -2 \Doa{-2} -3\Doa{-1}{}^2\Bigg]\cs\; .
\end{equation*}
Again it's convenient to define differential operators
\begin{eqnarray*}\fl
\Doa{n}:=\ve^{n+1}\ell_{n+1}^\l \p_\ve-\ve^n \l(1-\l)\ell_n^\l \p_\l
\qquad \textrm{with} \quad
\ell_n^\l :=(1-\l)^n-(-\l)^n
\end{eqnarray*}
and single variable counterpart
\begin{equation*}
\Dela{n}{m}:=m\, \ell_{n+1}^\l-\l(1-\l) \ell_n^\l \p_\l,
\end{equation*}
such that
\begin{eqnarray*}
\Doa{n} \ve^m F(\l) L_{\{k\}}\keto{\op}
=\ve^{n+m} \Dela{n}{m}F(\l) L_{\{k\}}\keto{\op}\\
\Doa{n} \ve^m G(\l) L_{\{k\}}\xs
=\ve^{n+m} \Dela{n}{m}G(\l) L_{\{k\}}\xs+ \ell_{n+1}^\l G(\l)L_{\{k\}}\keto{\pop}\; .
\end{eqnarray*}
Acting on $\cs$ with this null state operator produces the single novel constraint:
\begin{equation}
0=\Eq_{\l \ve} :=\left(3\Dela{-1}{0} \Dela{-1}{1}+2\Dela{-2}{1}\right)C_0^\l-\frac{2k_2(1-2\l)}{[\l(1-\l)]^{2/3}}-\frac{2}{\l} A^\l\; .
\end{equation}
 
Finally, the $\op(\ve)$ null state condition is expressed by letting $\{x,y\}=\{-\ve(1-\l),- \ve \}$:
\begin{equation*}
0=\Bigg[ 2 \left( \sum_{s=-1}^\infty \ve^s L_{-(s+2)}\right) - 2 \Dob{-2} -3\Dob{-1}{}^2\Bigg]\cs\; .
\end{equation*}
We define differential operators
\begin{equation*}
\Dob{n}:=(-\ve)^n\left( \ve \p_\ve+(1-\l)\left(1-(1-\l)^n \right) \p_\l \right)
\; ,
\end{equation*}
with single variable counterparts
\begin{equation*}
\Delb{n}{m}:=(-1)^n\left(m +(1-\l)\left(1-(1-\l)^n \right)  \p_\l\right)
\end{equation*}
such that
\begin{eqnarray*}
\Dob{n} \ve^m F(\l) L_{\{k\}}\keto{\op}
=\ve^{n+m} \Delb{n}{m}F(\l)  L_{\{k\}} \keto{\op}\\
\Dob{n} \ve^m \left( G(\l) \xs \right)
=\ve^{n+m}\Delb{n}{m}G(\l) \xs
+(-1)^n G(\l) \keto{\pop}\; .
\end{eqnarray*}
Acting on $\cs$ with this null state operator produces another new constraint:
\be
0=\Eq_{\ve} :=\left(3\Delb{-1}{0} \Delb{-1}{1}+2\Delb{-2}{1}\right)C_0^\l
+ \frac{k_2(3-\l)\l}{[\l(1-\l)]^{2/3}}
-2 A^\l\; .
\ee

Each $\op$ operator therefore allows us to derive one new second order non-homogeneous constraint equation. The sum
\begin{eqnarray}
0
&=\l\, \Eq_0-\l(1-\l) \Eq_{\l \ve}+(1-\l) \Eq_\ve\nn
&=3 k_2 (1+2 \a)(\l(1-\l))^{1/3}
\end{eqnarray}
shows that together the three null-state conditions imply that $\a=-1/2$.
This conclusion may seem questionable if $k_2=0$.  But in that case $\keto{\poplog}$ does not appear in $\cs$, so the apparent ambiguity occurs because $\a$ does not exist.

If we now set all occurrences of $\a$ to the appropriate value then we can determine $C_0^\l$ by taking
\begin{eqnarray*} \fl
0&=\frac{\Eq_0 -(1-\l)^2\Eq_{\l \ve}}{6\l^{1/3}(1-\l)^{4/3}}
=k_1 \frac{(1-\l)^{2/3}}{3 \l^{4/3}}
-\frac{k_2}{4}{}_2F_1(2/3,1;7/3;\l)
+\p_\l\left(\frac{C_0^\l}{[\l(1-\l)]^{1/3}} \right)\; .
\end{eqnarray*}
Solving this differential equation gives us
\begin{eqnarray} \label{equ:C0_func}
C_0^\l= c_1 \F^\l_1+c_2 \F^\l_2+c_3 \F^\l_3\; ,
\end{eqnarray}
 where we've labeled the functions
\begin{eqnarray*}
\F^\l_1 &=1-\frac{\l^2}{5}{}_2F_1\left(1,4/3;8/3;\l \right)\\
\F^\l_2 &=\frac{3\, \Gamma(5/3)}{ 4\, \Gamma(1/3)^2}\l^{4/3}(1-\l)^{1/3}{}_3F_2\left( 2/3,1,1;2,7/3;\l\right)\\
\F^\l_3 &=\frac{\Gamma(5/3)^2}{ \Gamma(7/3)} \l^{1/3}(1-\l)^{1/3}\; ,
\end{eqnarray*}
and changed the normalization of the functions to simplify the crossing relations in (\ref{equ: Xing rels}).  We make the same change to the normalization of the other terms in the OPE with the final result
\begin{equation}\fl
\cs = \left(c_1+c_2 \Pi^\l_h\right) \keto{\op}+c_2\, \ve\,  \frac{\sqrt{3}}{\pi} \F^\l_3 \xs
+\ve\, \big(c_1 \F^\l_1+c_2 \F^\l_2+c_3 \F^\l_3 \big) \keto{\pop}
+\mathrm{O}(\ve^2)\; .
\end{equation}

The three constants $c_1,c_2$ and $c_3$ are the only free parameters that remain.  These parameters appear independently at $\mathrm{O}(\ve)$ we can uniquely identify the conformal blocks by considering only terms at this order.  To identify the internal fusion channels we isolate the individual Frobenius series in $\l$ and $(1-\l)$ that make up the functions in (\ref{equ:C0_func}).

If we consider expansions in the variable $\l$, then two functions have leading exponent zero: $1$ and $\F^\l_1$.  So these functions indicate the presence of the identity in the $\l \to 0$ fusion.  We isolate this conformal channel:
\begin{equation*}
\cs^0_{1}:=\op(\ve) \left\{\op(\l \ve) \keto{\op} \right\}_{1}
= \keto{\op}+\ve\, \F^\l_1\, \keto{\pop}+\mathrm{O}(\ve^2)\; .
\end{equation*} 
The choice $c_1=1$ guarantees that the weight of the two point function is $1$.

Functions $\Pi^\l_h$, $\F^\l_2$ and $\F^\l_3$ all have series expansions with leading exponent $h_{1,3}=1/3$ indicating the presence of $\opa$ in the $\l \to 0$ fusion.  Isolating this fusion channel gives
\begin{eqnarray*} \fl
\cs^0_{\opa}&:=\op(\ve) \left\{\op(\l \ve) \keto{\op} \right\}_{\opa}  \\ \fl
&\;= \Pi_h^\l \keto{\op}
+\ve\,  \frac{\sqrt{3}}{\pi} \F^\l_3 \keto{\poplog}+\ve\left(\F^\l_2
+\frac{\sqrt{3}}{\pi}  \log (q_3 \ve) \F^\l_3\right)\keto{\pop}
+\mathrm{O}(\ve^2)\; .
\end{eqnarray*}
We are compelled to pick $c_2=1$ in order to reproduce the horizontal crossing probability when the result is applied within the four-point function.  The remaining parameter $c_3$ is absorbed into the logarithm as $q_3$ and remains unfixed as a result of the $\poplog \to \poplog+a\,  \pop$, symmetry of the operator algebra.

These two expressions exhaust the $\l \to 0$ internal blocks.  We can determine the two $\l \to 1$ internal blocks in a similar fashion using the crossing relations
\be
\label{equ: Xing rels}
\begin{array}{r @{\; =\; } l}
\Pi_h^{1-\l} & 1-\Pi_h^\l\\
\F^{1-\l}_1 & 1-\F^\l_1+2 \F^\l_3\\
\F^{1-\l}_2 & 1-\Pi_h^\l-\F^\l_1+\F^\l_2+\F^\l_3\\
\F^{1-\l}_3 & \F^\l_3
\end{array}
\ee 
that follow from standard relations for hypergeometric functions.

This fixes the internal $\l \to 1$  identity block,
\be \nonumber
 \cs^1_{1}
= \keto{\op}+\ve\, (\F^\l_1-2\F^\l_3) \keto{\pop}+\mathrm{O}(\ve^2)\; ,
\ee 
uniquely.  But the issue of fixing the remaining $\opa$ block is subtle due to the ambiguous $q_3$ term.

We begin with an expression for the $\l \to 0$ internal block as in (\ref{modansatz}) and letting $\{\ve, \l \} \to \{-\ve,1-\l\}$.  Then by shifting the positions of the operators by $\ve$ we recover the expression
\begin{eqnarray*} \fl
 \cs^1_{\opa}&:=\left\{ \op(\ve) \op(\l \ve) \right\}_{\opa}  \keto{\op}
=\E^{\ve L_{-1}} \op(-\ve) \left\{\op(-(1-\l) \ve) \keto{\op} \right\}_{\opa}
\\ \fl
&\;=\Pi_h^{1-\l} \keto{\op}
-\ve\,  \frac{\sqrt{3}}{\pi} \F^{1-\l}_3 \left[\keto{\poplog}+  \log (-q_3 \ve) \keto{\pop}\right]
+\ve\left[\Pi_h^{1-\l}\!\!-\F^{1-\l}_2
\right]\keto{\pop}
+\mathrm{O}(\ve^2)\; .
\end{eqnarray*}
for the corresponding $\l \to 1$ internal block.

By determining the conformal blocks in this way we are trying to make a consistent choice of $q_3$. 
But an ambiguity remains that stems from the minus sign in the logarithm.  Interestingly this has a simple interpretation in terms of vertex operators.
If the position of the operators is changed by rigid rotation through an angle $\pm \pi$, then $\ve \to \E^{\pm \pi \I} \ve$.  These choices of branch cut allow us to collect the four internal blocks into the single relation
\be
0=\cs^0_{1,3}+ \cs^1_{1,3}+\E^{\pm 2 \pi \I/3} \cs^0_{1,1}\E^{\pm 4\pi \I/3} \cs^1_{1,1}\; .
\ee
Comparing this relation to figure \eref{equ:4pt_CB_Relation} we see that this relation is precisely the relation Dotsenko an Fateev observed by manipluating screening charges \cite{Dotsenko1984}.

\section*{The Bibliography}

\end{document}